\documentclass[epj,preprint]{svjourblank}
\usepackage[centertags]{amsmath}
\usepackage{amsfonts}
\usepackage{amssymb}
\usepackage{fancyhdr}
\usepackage{amsmath}
\usepackage{graphics}

\newcommand{\D}[2]{\frac {\partial #1}{\partial #2}}
\renewcommand{\vec}[1]{\mbox{\boldmath$#1$}}
\newcommand{\lmode}{\ell}
\newcommand{\erg}{\,\mathrm{erg}}
\newcommand{\s}{\,\mathrm{s}}

\newcommand{\cm}{\,\mathrm{cm}}
\newcommand{\m}{\,\mathrm{m}}

\begin{document}
\title{Dynamics of shape fluctuations of quasi-spherical vesicles revisited} 
                   
\author{Ling Miao\thanks{\emph{e-mail:} miao@fysik.sdu.dk}%
 \and Michael A. Lomholt\thanks{\emph{e-mail:} mlomholt@fysik.sdu.dk} \and Jesper Kleis}

\institute{The MEMPHYS Center for Biomembrane Physics, Physics Department, University of Southern Denmark,\\ DK-5230 Odense M, Denmark}


\date{
}

\abstract{
In this paper, the dynamics of spontaneous shape fluctuations of a single, giant
quasi-spherical vesicle formed of a single lipid species is revisited theoretically.
A coherent physical theory for the dynamics is developed based on a number of
fundamental principles and considerations and a systematic formulation of the theory
is also established.  From the systematic theoretical formulation, an analytical
description of the dynamics of shape fluctuations of quasi-spherical vesicles
is derived.  In particular, in developing the theory we have made a new interpretation
of some of the phenomenological constants in a canonical continuum description of fluid
lipid-bilayer membranes and shown the consequences of this new interpretation in terms
of the characteristics of the dynamics of vesicle shape fluctuations.  Moreover, we have
used the systematic formulation of our theory as a framework against which we have discussed
the previously existing theories and their discrepancies.  Finally, we have made a systematic
prediction about the system-dependent characteristics of the relaxation dynamics of shape
fluctuations of quasi-spherical vesicles with a view of experimental studies of the phenomenon
and also discussed, based on our theory, a recently published experimental work on the topic.
}
\maketitle

\section{Introduction}
Dynamics of fluid membranes was already observed a couple of centuries ago,
as ``flickering" seen under optical microscope of resting red blood cells.
The first physical study of the ``flickering" phenomenon of red blood cells
was made, however, only in 1975 by Brochard and Lennon \cite{brochard75}.  They
presented the first theory for the phenomenon, where they employed the concept
of bending elasticity due to Helfrich \cite{helfrich73} for describing the 
conformational flexibility of fluid membranes and associated the ``flickering"
with the dynamics of equilibrium shape fluctuations of red-cell
membranes that are thermally driven due to the conformational flexibility
of the membranes.  This association has since become one of the essential
ideas in theories for dynamics of fluid membranes.

During the last three decades, unilamellar lipid-bilayer vesicles reaching
micrometer sizes that can be routinely prepared in the laboratory have 
provided one of the most important classes of model systems for biological
cell membranes.  They allow for well-defined and quantitative physical 
investigations of the conformational behaviour of fluid membranes.  The 
relevant physics governing the {\em static} equilibrium shapes of giant,
unilamellar vesicles has been fully understood, as the result of a series
of both theoretical and experimental studies \cite{reviews,berndl,ling94,dobereiner}.
The static properties of equilibrium shape fluctuations have also been
investigated \cite{milner87,faucon89,mitov92}.  The understanding
of the dynamics of equilibrium shape fluctuations of the vesicles has not,
however, reached the same stage.  This is largely the consequence of the
fact that the dynamics of fluid membranes involves many more physical
factors than the statics.  The static properties of a fluid membrane in
thermodynamic equilibrium are entirely determined by thermodynamic energetics
of the membrane alone.  In contrast, the dynamics of the membrane is not only
related to the thermodynamic energetics, but also depends on the dynamics
of the bulk aqueous fluid in which the membrane is suspended: motion of
the membrane induces motion in the bulk fluid, which in turn exerts hydrodynamic
force on the membrane.  Furthermore, any intrinsic dissipation mechanisms of the
membrane will in principle play a role in the dynamics.   

Many studies, both experimental and theoretical, have actually focused on
the dynamics of fluid membranes.  The earlier work of Schneider {\it et al.}
\cite{schneider84} investigated experimentally by use of fluorescence microscopy
the fluctuation spectrum of giant, unilamellar and quasi-spherical vesicles.
In the theoretical interpretation of the experimental data, Schneider {\it et al.},
and later, Milner and Safran \cite{milner87}, treated the energetics of the
membrane as that of a {\em single} and {\em incompressible} fluid surface with
the Helfrich bending rigidity.  Moreover, they neglected any intrinsic 
dissipation of the membrane.  Their theory predicted a single relaxation
rate for each spherical-harmonic mode of the vesicle shape fluctuations.

The earlier notion that a lipid-bilayer membrane could be considered as a
single and incompressible fluid surface soon had to be modified.  The
relevance of both the {\em bilayer} architecture of the membrane and the
{\em compressibility} of the constituting monolayers was
first appreciated and demonstrated in the context of equilibrium shapes
of giant lipid-bilayer vesicles \cite{ling94,dobereiner,area_difference92,wiese92,waugh92}.
Evans and Yeung soon also pointed out the relevance of this new notion to the
conformational dynamics of giant vesicles \cite{evans92}.  Based on their analysis
of a tether-pulling experiment on giant vesicles, they proposed that the bilayer
structure of the membrane in fact allows the two monolayers to move relative to
each other laterally and that the relative motion is associated with dissipation
of energy phenomenologically described as {\em intermonolayer friction}.
Moreover, they suggested a coupling between bending of the membrane and
a local field of relative area dilation/compression of the monolayers, whose
relaxation in turn drives the relative intermonolayer motion.  In this theory,
therefore, the conformational dynamics of a fluid membrane would be coupled to
dynamics of the relative monolayer dilation/compression field.   

Following the idea put forward by Evans and Yeung \cite{evans92}, Seifert and
Langer formulated a theoretical description of the paradigmatic case of the
dynamics of equilibrium fluctuations of an almost planar bilayer membrane
embedded in a viscous aqueous medium \cite{udo93}.  They also discussed the relevance 
of their result to the experimental measurements made by spin-echo technique
on multilamellar membrane stacks.  No conclusive confirmation of the theory,
however, could be made due to the lack of direct experimental data.

Giant, quasi-spherical vesicles, whose fluctuations can be observed directly
by video microscopy, provide another type of systems, for which a theory for 
the dynamics can in principle be put to a quantitative test against experimental
data.  Yeung and Evans \cite{evans95}, and later, Bivas {\it et al.} \cite{bivas99},
considered such systems.  Each group presented a theoretical formulation of its
own for the dynamics of equilibrium fluctuations of quasi-spherical vesicles.
Although, qualitatively speaking, the two theories were similar in that
both used the ideas originally proposed by Evans and Yeung \cite{evans92}, they differed
both in their detailed descriptions of the thermodynamic energetics of the vesicles and
in their formulations of the equations of motion.  Consequently, the two theories yielded
different analytical results that characterize the dynamics.  It is, however, not
straightforward to justify either of the two theories.  Yeung and Evans employed a
description of the thermodynamic free energy of the vesicle membrane, that would be
consistent with the assumption that local density inhomogeneities in the monolayers
relaxed much faster than bending deformations.  But the assumption is inconsistent
with the basic starting point of their theory, which was to treat the dynamics of
bending and density inhomogeneities on the same footing.  Bivas {\it et al.}, while
proposing and using a phenomenological model for the thermodynamic free energy that
was consistent with their considerations of dynamics, adopted an {\it ad hoc} approach
to the formulation of equations of motion that ultimately governed the dynamics.  The
approach lacks the rigor that a systematic approach based on some fundamental
principles would otherwise have.  Its justification is, therefore, less than
transparent.  Moreover, a number of approximations were used in both theories,
for which the justifications were not clear either, due to the absence of a
systematic formalism.  It is thus our opinion that a physically consistent and
systematic theory for the dynamics of equilibrium fluctuations of quasi-spherical
vesicles is still needed.

Continuing experiments on giant, quasi-spherical vesicles by the use of phase
contrast video microscopy further underline the need for a well-defined theory.  
A recent analysis \cite{pott} of the experimental data obtained by the use of the
techniques indicates that the dynamics of equilibrium fluctuations
of giant vesicles involves more relaxation processes than the single process of
relaxation of bending predicted by the theory of Schneider {\it et al.} \cite{schneider84} 
and Milner and Safran \cite{milner87}.  The still improving time resolution
of the video-microscopy technique will also lead to a large amount of experimental
data of better quality.  Having a systematic and well-defined theory will
greatly facilitate the analysis and interpretation of the experimental data as
well as the quantitative test of the theory itself, thus furthering our understanding
of the phenomenon. 

The purpose of this paper is, therefore, two-fold.  First, we will present, based on
a number of fundamental principles and considerations, a coherent physical theory, and
a systematic formulation of the theory for the dynamics of equilibrium fluctuations of giant,
quasi-spherical vesicles formed of a single lipid species.  One fundamental consideration
underlying the theory is Onsager's regression hypothesis: the average regression of the
spontaneous thermal fluctuations in a macroscopic system obeys the sames laws as the
corresponding macroscopic  irreversible processes \cite{onsager}.  Following this hypothesis,
we may understand the dynamics of the thermal fluctuations in a giant vesicle by investigating
the near-equilibrium dynamics of the same system.  Our starting point for formulating the
near-equilibrium dynamics of a giant vesicle is then to describe the vesicle dynamics
as the hydrodynamics of two coupled systems of (quasi-)two-dimensional fluid with
time-dependent surface geometry.  This description of the surface hydrodynamics of
the membrane thus naturally involves both a thermodynamic description of the membrane
and a description of the hydrodynamics of the bulk fluids surrounding the vesicle from
within and without.  Formulated this way, the theory, including the various approximations
that will be made, appears conceptually transparent and systematic.  This feature not only
will make the comparison between the theory and the experiments, and in turn, any
necessary modifications easier, but will also lend the formalism readily to any
extension that deals with vesicle systems of more complexity, for example,
giant vesicles made either of two lipid species or of a lipid species with
a membrane protein species.  Secondly, we will present, as an application of the
general formalism, a specific derivation of analytical expressions for the regression
dynamics of shape fluctuations of quasi-spherical vesicles.  In particular we discuss
systematically the different approximations employed in the application.  Our results
are different from those derived by Yeung and Evans \cite{evans95} and by Bivas {\it et al.}
\cite{bivas99}, and we will discuss the differences.  To facilitate the comparison
of the theory with the experiments, we also present a systematic numerical
analysis of the analytical results.  

The plan of the presentation is as follows.  In Section 2, we introduce and
discuss a thermodynamic description of fluid lipid-bilayer membranes.  In 
Section 3, we present a systematic formulation of near-equilibrium dynamics of
fluid-bilayer vesicles, which can in principle be applied to vesicles with
arbitrary equilibrium shapes.  In Section 4, we apply the general formulation
described in Section 3 to systems of quasi-spherical vesicles.  Finally, in
Section 5, we discuss the implications that our results have for experimental
measurements on the dynamics of spontaneous shape fluctuations of quasi-spherical
vesicles.  We will also discuss the differences and connections between our results
and the theoretical results obtained by other groups previously \cite{evans95,bivas99}.
 
\section{Thermodynamic description of fluid lipid-bilayer membranes}
In this section, we will consider a vesicle formed of a closed fluid lipid-bilayer
membrane containing only a single lipid species.  The vesicle encloses a volume
of $V$, which can be considered fixed under typical experimental conditions.  
The two constituting monolayers of the vesicle membrane, the outer one labeled by
superscript ``+" and the inner one by ``$-$", contain $N^{+}$ and $N^{-}$ lipid molecules,
respectively.  As has been recognized \cite{ling94}, $N^{+}$ and $N^{-}$ depend on the
vesicle formation process and may be different from each other.  A total area,
$A_{\mathrm{m}}$, that can be resolved with techniques of microscopic resolutions
or with equivalent means, can be defined for the whole membrane.  To a good approximation,
$A_{\mathrm{m}}$ may be treated, for most of the membrane systems studied, as a fixed
quantity that depends on $N^{+}$ and $N^{-}$.  When a vesicle is flaccid, in other
words, having a non-zero excess area $\Delta \equiv A_{\mathrm{m}}/4\pi R_{\mathrm v}^{2}-1$,
where $R_{\mathrm v} \equiv (3V/4\pi)^{1/3}$, it undergoes thermally driven
conformational fluctuations as a result of the membrane flexibility.  In the
case of giant vesicles that are typically investigated, whose sizes are in 
the micrometer range and whose bending rigidities are of the order of
$10^{-19}\;{\mathrm J}$, the fluctuations remain controlled.  Thus, the vesicle
we consider has a well-defined equilibrium shape, whose specific geometry depends on
$V$, $N^{+}$, $N^{-}$ and temperature $T$.  

We will present a phenomenological description of the thermodynamics
of a fluid lipid-bilayer membrane, which is consistent with our starting point that
the membrane is conformationally flexible and consists of two weakly coupled
compressible monolayers.  This description will form a necessary part of the
formulation of the membrane dynamics.

\subsection{{\sffamily Notations and relevant fields}}
To describe the conformational flexibility of a fluid membrane, we use the
basic notation of surface differential geometry.  At length scales larger
than molecular lengths, the varying conformation of a fluid membrane suspended in
a bulk aqueous fluid can be approximated by a two-dimensional surface embedded in the
three-dimensional space and with a time-dependent geometry.  Such a surface can be 
represented by a three-dimensional vector, $\vec{R}(u^1,u^2,t)$, where $u^1$ and $u^2$
are the two coordinates parametrizing the two-dimensional internal space and $t$
represents time.  The derivatives of $\vec{R}(u^1,u^2,t)$ with respect
to the internal coordinates $u^\alpha\;(\alpha=1,2)$ define two local tangent vectors
to the surface: 
\begin{equation}
\label{eq1:geoTangent} 
\vec{t}_\alpha(u^1,u^2,t)\equiv
\frac{\partial \vec{R}(u^1,u^2,t)}{\partial u^\alpha}
\equiv \partial_{\alpha} \vec{R}(u^1,u^2,t), \; \; \alpha=1,2.
\end{equation}
From $\vec{t}_{1}$ and $\vec{t}_{2}$, a local unit vector
normal to the surface can be constructed:
\begin{equation}
\label{eq1:geoNormal} 
\vec{n} \equiv \frac{\vec{t}_{1} \times \vec{t}_{2}}
                 {|\vec{t}_{1} \times \vec{t}_{2}|}\;.
\end{equation}
The metric tensor $a_{\alpha\beta}$ of the surface is then given by
\begin{equation}
\label{eq1:geoMetric} 
a_{\alpha\beta}=\vec{t}_\alpha \cdot \vec{t}_\beta \;,
\end{equation}
which is a symmetric tensor with a non-zero determinant
$a \equiv \det(a_{\alpha\beta})$.  A local area element, $dA$, can be expressed
as
\begin{equation}
\label{eq1:geoAreaElement} 
dA=\sqrt{a}\, du^1 du^2.
\end{equation}
The inverse of the metric tensor, $a^{\alpha\beta}$, follows from 
\begin{equation}
a^{\alpha\beta} a_{\beta \gamma} = \delta^{\alpha}_{\gamma} \;,
\end{equation}
leading to the definitions of contravariant surface tangent vectors 
$\vec{t}^{\alpha} \equiv a^{\alpha \beta}\vec{t}_\beta$.  The rule of summation
over repeated Greek indices has been used in the above expressions, and will be
kept to in the rest of the paper.  Clearly, 
$\vec{t}^{\alpha} \cdot \vec{t}_\beta = \delta^{\alpha}_{\beta}$.  
Any space vector $\vec{w}$ tangent to the surface can be expressed as 
$\vec{w} = w_{\alpha}\vec{t}^{\alpha} = w^{\alpha}\vec{t}_{\alpha}$, where
$w_{\alpha} = \vec{w} \cdot \vec{t}_{\alpha}$ and 
$w^{\alpha} = \vec{w} \cdot \vec{t}^{\alpha} =a^{\alpha\beta}w_{\beta}$ are
the covariant and contravariant components of $\vec{w}$, respectively. 

A full description of the essential local properties of the surface requires
also the information contained in 
\begin{equation}
\partial_{\alpha}\partial_{\beta}\vec{R}=\Gamma^{\gamma}_{\alpha \beta}\partial_{\gamma}\vec{R}
                                +b_{\alpha \beta}\vec{n}\; ,
\end{equation} 
where $b_{\alpha\beta} \equiv \vec{n}\cdot \partial_\alpha \partial_\beta \vec{R}$
is the so-called ``curvature tensor", and $\Gamma^{\gamma}_{\alpha \beta}$ defines
the Christoffel symbol.  
From the curvature tensor both the mean curvature $H$ and the Gaussian
curvature $K$ can be obtained, respectively, 
\begin{equation}
\label{eq1:geoMean} 
H \equiv \frac{1}{2}a^{\alpha\beta}b_{\alpha\beta}
= \frac{1}{2} \left( \frac{1}{R_{1}} + \frac{1}{R_{2}} \right) \;,
\end{equation}
and 
\begin{equation}
\label{eq1:geoGauss} 
K \equiv \det \left( a^{\alpha\gamma}b_{\gamma\beta} \right)
= \frac{1}{R_{1}}\cdot \frac{1}{R_{2}} \;
\end{equation}
where $R_{1}$ and $R_{2}$ are the two principal radii of curvature.

By use of the Christoffel symbol $\Gamma^{\gamma}_{\alpha \beta}$, covariant
differentiations with respect to $u^{\alpha}$, denoted by $D_{\alpha}$, of surface scalars,
vectors and tensors can be performed.  For any surface scalar, $\rho(u^{1},u^{2})$,
$D_{\alpha}\rho(u^{1},u^{2}) = \partial_{\alpha} \rho(u^{1},u^{2})$, and for a surface
vector represented by its contravariant components, $w^{\alpha}$, 
$D_{\alpha}w^{\gamma} = \partial_{\alpha}w^{\gamma}+ 
                        \Gamma^{\gamma}_{\alpha \beta}w^{\beta}$.
                        
In order to fully characterize the physical state of a flexible and compressible
fluid lipid-bilayer membrane formed of a single lipid species, two more fields
are needed in addition to $\vec{R}(u^1,u^2,t)$: $\rho^{+}(u^1,u^2,t)$ and 
$\rho^{-}(u^1,u^2,t)$, representing the local surface (number) densities of the
outer and inner monolayer, respectively, and defined with respect to the surface
described by $\vec{R}(u^1,u^2,t)$.      

\subsection{{\sffamily Thermodynamic free energy}}
Imagine a vesicle suspended in a bulk fluid at temperature $T$.  Its equilibrium
(relaxed) state is then characterized both by an average shape $\vec{R}_{0}(u^1,u^2)$
and by average monolayer densities $\rho_{0}^{+}(u^1,u^2) \equiv N^{+}/A_{0}$ and 
$\rho_{0}^{-}(u^1,u^2) \equiv N^{-}/A_{0}$ \footnote{Strictly speaking,
$\rho_{0}^{+}(u^1,u^2)$ and $\rho_{0}^{-}(u^1,u^2)$ may not be constant for non-spherical
shapes.}, where $A_{0}$ is the area defined by the average shape.  For later use, we
define 
\[
N_{\Sigma}\equiv \frac{N^{+}+N^{-}}{2} \;, \quad 
N_{\Delta}\equiv \frac{N^{+}-N^{-}}{2} \;.
\]
Assume now that the vesicle is in a near-equilibrium state with an effective shape
described by $\vec{R}(u^1,u^2,t)$ and with density fields $\rho^{+}(u^1,u^2,t)$ and
$\rho^{-}(u^1,u^2,t)$ describing respective inhomogeneities in the two monolayers.
We associate the following thermodynamic free energy to the near-equilibrium state
of the vesicle:
\begin{eqnarray}
\label{model} 
F =&& 
    \int dA \; \frac{k_{\mathrm{eff}}}{2} \; \left( \frac{\rho^{+}}{\rho_0}-1 \right)^2
    + \int dA \; \frac{k_{\mathrm{eff}}}{2} \; \left( \frac{\rho^{-}}{\rho_0}-1 \right)^2
    \nonumber \\
    &+&\int dA \; \gamma (\rho^{+} -\rho_0) + \int dA \; \gamma (\rho^{-} -\rho_0)
      +\sigma \int dA  
    \nonumber \\
    &+&\int dA \; \frac{\kappa_{\mathrm{eff}}}{2} (2H)^2  
    + \int dA \; \lambda H (\frac{\rho^{+}}{\rho_0} - \frac{\rho^{-}}{\rho_0}) \; \;.
\end{eqnarray}
The five terms in the first two lines may be viewed as the result of an expansion of
a local free energy function,
$F_{T,\Delta}(N^{+}_{\Delta},N^{-}_{\Delta},A_{\Delta})= A_{\Delta}f(T,\rho^{+},\rho^{-})$,
about a particular reference value, $\rho_0$, of the monolayer surface densities.  
In other words, the first four terms describe the free-energy cost associated with
deviations in the monolayer surface densities from the reference value and the 
last of the five terms represents the free energy required to change the area of the
effective shape of the vesicle at constant density, $\rho_{0}$.  Consequently,
the constant parameters, $k_{\mathrm{eff}}$, $\gamma$, and $\sigma$, all depend on
the chosen reference state.  $k_{\mathrm{eff}}$, in particular, is the monolayer
compressibility modulus corresponding to the chosen $\rho_0$.  The first term in
the last line of Eq.(\ref{model}) represents the free-energy cost associated with 
bending of the membrane, and the second term takes into account a possible coupling
between the bending of the membrane and the difference between local monolayer surface
densities: a non-zero $(\rho^{+} - \rho^{-})$ may be viewed as amounting to a dynamic
spontaneous curvature \cite{udo93,udo97,per98}.

Eq.(\ref{model}) can be rewritten as
\begin{eqnarray}
\label{model2a} 
F &=& 
    \int dA \; \frac{k_{\mathrm{eff}}}{2} \; \left( \frac{\rho^{+}}{\rho_0}-1 \right)^2
    + \int dA \; \frac{k_{\mathrm{eff}}}{2} \; \left( \frac{\rho^{-}}{\rho_0}-1 \right)^2
\nonumber\\
&&    +\int dA \; \gamma \rho^{+}  + \int dA \; \gamma \rho^{-} 
    +\int dA \; \frac{\kappa_{\mathrm{eff}}}{2} (2H)^2  
\nonumber \\
&& + \int dA \; \lambda H (\frac{\rho^{+}}{\rho_0} - \frac{\rho^{-}}{\rho_0}) 
   +\sigma_{0} \int dA \; \;,
\end{eqnarray}
where $\sigma_{0} \equiv \sigma - 2\gamma \rho_0$.  The terms
$\int dA \; \gamma \rho^{+}$ and $\int dA \; \gamma \rho^{-}$ will be dropped henceforth
since we will only consider situations where the number of lipid molecules in each
monolayer is conserved.  

A couple of important remarks on the physical interpretation of the above free energy
are due here.  First, since our theory is intended to describe the dynamics of
vesicle conformation that can be resolved at optical or {\em mesoscopic} length scales, 
the free energy should be understood as an effective Hamiltonian resulted from
an appropriate coarse-graining procedure which integrates over fluctuations at
suboptical scales \cite{cai95,peliti01}. The work presented in Refs.~\cite{cai95,peliti01} makes it conceptually clear that the physical parameters describing the elastic
properties of the vesicle membrane at the mesoscopic length scales, namely,
$k_{\mathrm{eff}}$, $\kappa_{\mathrm{eff}}$, and $\lambda$, are ``renormalized"
parameters, different in principle from those parameters describing the elastic
properties of the membrane at {\em microscopic} length scales.  For the same reason,
the constant $\sigma_{0}$ conjugate to the ``apparent" membrane area under an optical 
resolution is an {\em effective membrane tension} \cite{peliti01}, and is distinct both
from the microscopic surface tension and from the mechanical frame tension
which is associated with the equilibrium shape and which can be accessed
by micromechanical measurements \cite{david91,rawicz}.   

This issue has not been addressed before in the previous works on membrane
and vesicle dynamics \cite{udo93,evans95,bivas99}.  And, one might argue that, when
fluctuations are sufficiently small to allow a Gaussian (linear) theory of the
fluctuations, the distinction between the effective elastic parameters
at the mesoscopic length scales and the elastic parameters at microscopic
length scales would only be conceptual rather than quantitative, as the conventional
wisdom on linear theories of fluctuations may lead one to believe.  However, the work
reported in Refs.~\cite{cai95,peliti01} clearly shows that care must be taken when
we apply the conventional wisdom to fluid membrane systems.  In particular, if we
consider a vesicle whose membrane is nearly incompressible at microscopic length
scales and which is in either a {\em floppy} state or a {\em entropic-tense} state
\cite{peliti01}, the effective area compressibility modulus $k_{\mathrm{eff}}$ of
the vesicle membrane at mesoscopic length scales originates almost entirely from
integrating over small fluctuations at suboptical length scales and is smaller
than the microscopic area compressibility modulus.  We will show in later 
discussions that our new interpretation of the physical parameters in the
model free energy bears consequences for the qualitative as well as quantitative
characterization of vesicle dynamics.

The second remark concerns the validity of the expansion form of the free
energy, and relatedly, the choice of the reference state, represented by $\rho_0$.
As discussed in Ref.~\cite{peliti01}, the effective free energy associated
with optical length scales involves a {\em nonlinear} area elasticity.  Therefore,
the expansion, which only contains {\em linear} area elasticity, is valid in principle
only when deviations from the chosen reference state $\rho_0$ are small.  Thus, 
for our considerations of near-equilibrium states, a natural choice of $\rho_0$ 
is
\begin{equation}
\rho_0 =N_{\Sigma}/A_{0} \;,
\end{equation}

Finally, with definitions $\phi^{\pm} \equiv \rho^{\pm}/\rho_{0}-1$, Eq.(\ref{model2a})
can be rewritten as 
\begin{eqnarray}
\label{model2b} 
F &=& 
    \int dA \; \frac{k_{\mathrm{eff}}}{2} (\phi^{+})^2
    + \int dA \; \frac{k_{\mathrm{eff}}}{2} (\phi^{-})^2 
    \nonumber \\
    &&+\int dA \; \frac{\kappa_{\mathrm{eff}}}{2} (2H)^2  
    + \int dA \; \lambda H (\phi^{+} -\phi^{-} ) \nonumber \\
    &&+\;\sigma_{0} \int dA \; \;.
\end{eqnarray}

\section{{\sffamily A systematic and general formulation of vesicle dynamics }}
In this section, we will present a formulation of vesicle dynamics.  The formulation
is both systematic, relying on various basic concepts and notions from condensed
matter physics, and general, in that it can in principle be applied to vesicles
with arbitrary geometry. 

\subsection{{\sffamily Surface hydrodynamics of the membrane: Equations of motion}}
Following our discussion of vesicle dynamics in {\bf {\sffamily Introduction}},
we begin with a hydrodynamic description of the surface flows within the two
fluid monolayers constituting the vesicle membrane.  The description treats the two
monolayers as two individual, but weakly coupled, systems of quasi-two-dimensional
fluids; it also takes into account the fact that each monolayer interacts with
its corresponding bulk fluid environment.  Moreover, it is assumed in the description
that the motions within the two monolayers take place under isothermal conditions,
leaving to be considered for each monolayer only two intrinsic hydrodynamic fields,
the density field and the surface velocity field.  Consequently, the 
surface hydrodynamics of each monolayer is principally governed by two equations
of motion: one states the law of mass (particle) conservation,
and the other expresses the generalized law of momentum conservation.

To facilitate the presentation, we first introduce some additional notations
related to the dynamics.  Let's consider a fluid surface described by
$\vec{R}(u^1,u^2,t)$, and let the trajectory of a particular material particle
be represented by $u^{\alpha}= u^{\alpha}(\xi,t), \alpha =1,2$ 
where $\xi$ labels the material particle.
The covariant components of the intrinsic velocity of the particle are then
defined by
\begin{equation}
W^{\alpha} \equiv \frac{du^{\alpha}}{dt} 
\equiv \left. \frac{\partial u^{\alpha}(\xi,t)}{\partial t}
\right|_{\xi} \;.
\end{equation} 
The material velocity of the particle in the three dimensional embedding
space can be obtained easily:
\begin{eqnarray}
\vec{U}& \equiv &\left. \frac{d\vec{R}(u^1,u^2,t)}{dt} \right|_{\xi}
= W^{\alpha} \vec{t}_{\alpha} 
+ \left. \frac{\partial \vec{R}(u^1,u^2,t)}{\partial t}\right|_{u}\nonumber\\
&\equiv & W^{\alpha} \vec{t}_{\alpha} + \partial_{t} \vec{R} \;.
\end{eqnarray}
$\vec{U}$ may also be expressed as a decomposition into components tangential
and normal to the surface,
\begin{equation}
\vec{U} =\vec{U}_{\mathrm{t}} + U_{\mathrm{n}} \vec{n} \;,
\end{equation}
where $U_{\mathrm{n}} \equiv \vec{U}\cdot \vec{n} $ and
$\vec{U}_{\mathrm{t}} = U^{\alpha} \vec{t}_{\alpha}$.  $U^{\alpha}$ is in general not
equal to $W^{\alpha}$ as a result of the time-varying geometry, but is related to 
$W^{\alpha}$ as follows,
\begin{equation}
\label{decomposition}
U^{\alpha} = W^{\alpha} + a^{\alpha \beta} \vec{t}_{\beta}\cdot (\partial_{t} \vec{R}) \;. 
\end{equation}

For any function $f(u^1,u^2,t)$ defined on the fluid surface, a useful time
derivative may be defined:
\begin{equation}
D_{t}f(u^1,u^2,t) \equiv a^{-1/2} \partial_{t} (a^{1/2}f)
= \partial_{t} f + f \frac{\partial_{t} a}{2a} \;.
\end{equation}

\subsubsection{{\sffamily Equations of motion}}
We can now write down the four principal equations of motion, two for each monolayer,
that govern the dynamics of a fluid lipid-bilayer vesicle.  The formalistic derivations
of the equations are given elsewhere \cite{cai95,rutherford}.
Given the monolayer surface density fields, $\rho^{\pm}(u^1,u^2,t)$, and 
the intrinsic velocity fields, $W_{\pm}^{\alpha}(u^1,u^2,t)$, we can define
intrinsic particle fluxes, $j_{\pm}^{\alpha} \equiv \rho^{\pm}W_{\pm}^{\alpha}$.
The subscripts ``+" and ``$-$", same as the superscripts, refer to the outer and the
inner monolayer of the vesicle, respectively.  The law of mass (particle) conservation
for each monolayer can then be expressed in covariant form as follows,
\begin{equation}
\label{surfacecontinuity}
D_{t}\rho^{\pm} + D_{\alpha} j_{\pm}^{\alpha}  = 0 \;,
\end{equation}
where $D_{t}$ and $D_{\alpha}$ have been defined previously.

The generalized law of momentum conservation can also be written in a similar way.
For each monolayer, a three-dimensional vector representing the surface density of
monolayer momentum can be defined by
$\vec{J}_{\pm} \equiv m \rho^{\pm} \vec{U}^{\pm} 
 = m \rho^{\pm} ( W_{\pm}^{\alpha} \vec{t}_{\alpha} + \partial_{t} \vec{R})$,
where $m$ is the molecular mass of the constituting lipid.
The equations of motion governing the changes of $\vec{J}_{\pm}$ take the following
form:
\begin{equation}
\label{surfacedynamics}
D_{t}\vec{J}_{\pm} + D_{\alpha} (j_{\pm}^{\alpha} \vec{J}_{\pm}) = \vec{f}^{\pm}
\;,
\end{equation}
where $j_{\pm}^{\alpha} \vec{J}_{\pm}$ are the covariant momentum fluxes.  
$\vec{f}^{\pm}$ represent the total forces per unit area acting on the monolayers.

\subsection{{\sffamily Forces acting on monolayer surfaces}} 
In principle $\vec{f}^{\pm}$ should contain several forces of different origins: 
any dissipative forces associated with the intrinsic surface shear and dilational
viscosities of the monolayers; $\vec{f}_{\rm{m}}^{\pm}$, forces which the two
monolayers exert on each other;  $\vec{f}_{\mathrm{rs}}^{\pm}$, the mechanical
(restoring) forces associated with the free energy Eq.(\ref{model2b}); and, finally,
$\vec{T}^{\pm}$, the hydrodynamic forces exerted on the monolayers by their
corresponding bulk fluids.  In the following, we discuss each of the
four types of forces in turn. 

In our theory of vesicle dynamics, we have neglected the intrinsic dissipative
forces within each individual monolayer, based on the evidence and arguments given
in Ref. \cite{evans95}.  The intermonolayer forces, $\vec{f}_{\rm{m}}^{\pm}$, may be
modelled as 
\begin{equation}
\vec{f}_{\rm m}^{\pm} 
=  \mp b (\vec{U}^{+}_{\mathrm{t}} - \vec{U}^{-}_{\mathrm{t}} )+ \vec{f}_{\rm n}^{\pm} \; ,
\end{equation}
where
\begin{equation}
\vec{f}_{\rm b}^{\pm} 
\equiv \mp b (\vec{U}^{+}_{\mathrm{t}} - \vec{U}^{-}_{\mathrm{t}} ) \; ,
\end{equation}
phenomenologically describes dissipation associated with the relative motion between
the monolayers.  Thus, $b$ may be called the ``intermonolayer friction coefficient." 
This description was originally proposed by Evans {\it et al.} \cite{evans92}.  
$\vec{f}_{\rm n}^{\pm}$, with $\vec{f}_{\rm n}^{+} = -\vec{f}_{\rm n}^{-}$, represent
the normal components of the intermonolayer forces, which constrain the motions
of the two monolayers in the normal direction.  

\subsubsection{{\sffamily Mechanical restoring force}} 
In deriving from the free energy given in Eq.(\ref{model2b}) the mechanical restoring force acting on each individual monolayer, we may imagine the following infinitesimal virtual variation in the geometry of each monolayer:
\begin{equation}
\vec{R}(u^1,u^2,t) \rightarrow \vec{R}(u^1,u^2,t) + \delta \vec{R}(u^1,u^2) \;,
\end{equation}
and define the restoring forces $\vec{f}_{\mathrm{rs}}^{\pm}$ through
\begin{equation}
\label{variation}       
\delta F^{\pm} = \oint dA \; (-\vec{f}_{\mathrm{rs}}^{\pm}) \cdot \delta \vec{R}
 \; .
\end{equation} 
where
$\delta F^{+}$ and $\delta F^{-}$ are the variations in the free energies of the two monolayers $F^\pm$ induced by the shape variation $\delta \vec{R}$.

Two issues are implicit in Eq.(\ref{variation}), which need to be addressed clearly.
The first is the issue of how to evaluate $F^{\pm}$.  This issue has not
been discussed at all in the current literature on dynamics of lipid-bilayer membranes.  
Our proposal is to divide $F$ into two ``monolayer parts", $F= F^{+} + F^{-}$,
where
\begin{eqnarray}
F^{\pm} &=& 
 \oint dA \; \frac{k_{\mathrm{eff}}}{2} \cdot \left( \phi^{\pm} \right)^2
 \pm \oint dA \; \lambda H \phi^{\pm}\nonumber\\
&& + \frac{1}{2} \left\{ \oint dA \; \frac{\kappa_{\mathrm{eff}}}{2} (2H)^2 
                  + \sigma_0 \oint dA \; \right\}  \;   
\end{eqnarray}
is associated with a single monolayer, and then to derive the corresponding restoring
force from the variation of the monolayer free energy.  This definition is consistent
with the bilayer-composite structure of the vesicle membrane. 

The second is the issue of how to deal with the variations in the monolayer density
fields $\rho^{+}(u^1,u^2,t)$ and $\rho^{-}(u^1,u^2,t)$ in their relations to the
shape variations of the monolayer surfaces.  Our proposal \footnote{Our proposal
appears to be the same as that used in Ref.~\cite{cai95}.} concerning this issue
is that the number of molecules associated with any local area element $dA$
of a monolayer should be conserved under the shape variation of the monolayer.
In other words, 
\begin{equation}
\delta (dA \; \rho^{\pm}) = 0 \; .  
\end{equation} 
The reason for this lies in the basic thermodynamic consideration that the variation
in the free energy should be equal to the {\em mechanical} work done against the 
restoring forces.  Briefly, in considering the hydrodynamics of the membrane surface,
we assume that the thermodynamics of a local monolayer element with area $dA$ and 
$N_{\Delta}^{\pm}$ number of molecules is described by the local free energy
\begin{eqnarray}
&F_{\Delta}^{\pm}(T,N_{\Delta}^{\pm},dA(\vec{R}),\vec{R})
= dA \; \left\{ \displaystyle{\frac{k_{\mathrm{eff}}}{2}} \cdot \left( \phi^{\pm} \right)^2
     + \frac{\sigma_0}{2} \right\}  \nonumber\\
&              + dA \; \left\{ \displaystyle{\frac{\kappa_{\mathrm{eff}}}{4}} (2H)^2 \pm \lambda H \phi^{\pm} \right\}
  \;.              
\end{eqnarray}
Clearly, the change in the local free energy, which can be identified with mechanical
work, must be the change under constant $T$ and constant $N_{\Delta}^{\pm}$.  

The mathematical derivation of the restoring forces for a vesicle of arbitrary shape
is similar to that described by Jenkins \cite{jenkins77} and is somewhat lengthy.  
We will only state the final result here:    
\begin{align}
\begin{split}
\label{restoringforce} 
\vec{f}_\mathrm{rs}^{\pm}
=& \bigg\{ - \kappa_{\mathrm{eff}} \left[ H \left( 2H^2-2K \right) + \Delta H \right]- \pi^{\pm}(2H)\\
&\ \ \mp \lambda \left[ \frac{1}{2}\Delta\phi^{\pm} + (2H^2-K)\phi^\pm + 2 H^{2}
            \right]  \bigg\} \vec{n}   \\
 & - \vec{\nabla}_{\mathrm{t}} \pi^{\pm}
 \mp \lambda (1 + \phi^{\pm}) \vec{\nabla}_{\mathrm{t}} H \;,
\end{split}
\end{align}
where $\Delta\equiv(1/\sqrt a)\partial_{\alpha}(a^{\alpha\beta}\sqrt{a}\partial_{\beta})$
is the Laplace-Beltrami operator,
$\vec{\nabla}_{\mathrm{t}} \equiv \vec{t}_{\beta}a^{\alpha\beta}\partial_\alpha $ is
the gradient operator defined on the surface, and 
$\displaystyle \pi_{\pm} \equiv -\frac{\sigma_0}{2} + k_{\mathrm{eff}} \phi^{\pm}
                  +\frac{k_{\mathrm{eff}}}{2} \left( \phi^\pm \right)^2$ may be
considered as monolayer surface pressures under the condition of planar geometry.

\subsubsection{{\sffamily Hydrodynamic forces due to the bulk fluids }}
The hydrodynamic forces exerted on the monolayers by their corresponding bulk fluids
can be determined by solving the equations of motion for the two bulk fluids separated
by the membrane under appropriate boundary conditions. 
 
\noindent
{\small {\bf {\sffamily Equations of motion}}} \\
\noindent
Since the bulk solvents for lipid-bilayer vesicles are always aqueous,
it is a good approximation to consider the fluids as being incompressible.
Let the velocity and pressure fields in the two bulk fluids be represented by
$\vec{v}^{\mathrm{(o)}}(\vec{r},t)$, $\vec{v}^{\mathrm{(i)}}(\vec{r},t)$, and
$p^{\mathrm{(o)}}(\vec{r},t)$, $p^{\mathrm{(i)}}(\vec{r},t)$, respectively,
where the superscript ``o" refers to the bulk fluid outside the vesicle and in 
contact with the (+)-monolayer and the superscript ``i" refers to the bulk fluid
inside the vesicle and in contact with the ($-$)-monolayer.  It is then straightforward
to write down the hydrodynamic equations:
\begin{eqnarray}
\label{bulk}
&\nabla \cdot \vec{v}^{\mathrm{(a)}} = 0 \;, \\
&\rho_{\rm b} \left[ \frac{\partial\vec{v}^{\mathrm{(a)}}}{\partial t} 
                  + (\vec{v}^{\mathrm{(a)}} \cdot \nabla)\vec{v}^{\mathrm{(a)}} \right]
= -\nabla p^{\mathrm{(a)}} + \eta \nabla^{2} \vec{v}^{\mathrm{(a)}} \;,
\end{eqnarray}
where $\rho_{\rm b}$ and $\eta$ are the mass density and the shear viscosity of the bulk
fluids, respectively, and the superscript $\mathrm{a}=\mathrm{i},\mathrm{o}$.

\noindent
{\small {\bf {\sffamily Boundary conditions: kinematic matching }}} \\
\noindent
Each monolayer of the vesicle provides a boundary surface for the corresponding
bulk fluid.  Boundary conditions, which specify
$\vec{V}^{\mathrm{(a)}}(u^1,u^2,t) \equiv
 \vec{v}^{\mathrm{(a)}}(\vec{r}=\vec{R}(u^1,u^2,t),t)$, 
are required for completely determining solutions to the above equations.  Under very general circumstances, $\vec{V}^{\mathrm{(a)}}$ may be different from the flow velocity in the corresponding monolayer,
$\vec{U}^{\pm}(u^1,u^2,t)$.  In the absence of any practical evidence
for the general scenario, we choose to match the bulk kinematics at the boundary surfaces
with the monolayer surface kinematics, i.e.,
\begin{eqnarray}
\label{kinematic}
\vec{V}^{\mathrm{(o)}}(u^1,u^2,t) = \vec{U}^{+}(u^1,u^2,t)\;,\\
\vec{V}^{\mathrm{(i)}}(u^1,u^2,t) = \vec{U}^{-}(u^1,u^2,t)\;. 
\end{eqnarray}

\noindent
{\small {\bf {\sffamily Bulk hydrodynamic forces}}} \\
\noindent
Once the complete solutions to the bulk hydrodynamic equations are obtained,
the corresponding hydrodynamic stress tensors, $\mathsf{T}^{\mathrm{(a)}}$, are 
readily derived as,
\begin{equation}
\mathsf{T}^{\mathrm{(a)}}(\vec{r},t) = -p^{\mathrm{(a)}}(\vec{r},t) \mathbf{I} + 
\eta \left[ \nabla \vec{v}^{\mathrm{(a)}}(\vec{r},t) 
        + (\nabla \vec{v}^{\mathrm{(a)}}(\vec{r},t))^{\mathsf{T}}  \right],
\end{equation}
where $\mathbf{I}$ is the unit tensor, and the superscript ``$^{\mathsf{T}}$" indicates the
transpose of a tensor. The forces per unit area exerted on the two monolayers by the
corresponding bulk fluids can finally be evaluated as follows,
\begin{eqnarray}
&&\vec{T}^{+}(u^1,u^2,t) = \left. \mathsf{T}^{\mathrm{(o)}}(\vec{r},t) \right|_{\vec{r}=\vec{R}(u^1,u^2,t)} 
                      \cdot \vec{n}(u^1,u^2,t) \;,\nonumber\\
&&\vec{T}^{-}(u^1,u^2,t) = -\left. \mathsf{T}^{\mathrm{(i)}}(\vec{r},t) \right|_{\vec{r}=\vec{R}(u^1,u^2,t)} 
                      \cdot \vec{n}(u^1,u^2,t)  \;,\nonumber          
\end{eqnarray}
where $\vec{n}(u^1,u^2,t)$ indicates the local normal vector of the membrane surface that directs
towards the exterior of the vesicle.  

\section{{\sffamily Dynamics of shape fluctuations of quasi-spherical vesicles}}
Having formulated in the previous sections a systematic and general theory for
near-equilibrium dynamics of lipid-bilayer vesicles, we will in this section present
the application of the theory to systems of quasi-spherical vesicles.  By quasi-spherical
vesicles, we refer to those vesicles with very small excess areas such that their equilibrium
shapes are spherical.  Thus, both equilibrium shape fluctuations and non-equilibrium shape
deformations may be considered as small perturbations around the spherical equilibrium shapes.
Under such conditions, an analytical rendering of the theory becomes possible, provided that
certain approximations are made.

\subsection{{\sffamily Quasi-spherical vesicles: specific notations}}
For describing the surface geometry of a quasi-spherical vesicle in the embedding
three-dimensional space, the most convenient coordinate system is the spherical
coordinate system $(\theta,\varphi,r)$, with local unit basis vectors $\vec{e}_{\theta}$,
$\vec{e}_{\varphi}$, and $\vec{e}_{r}$.  The quasi-spherical geometry of the
membrane surface of the vesicle can be represented by
\begin{equation}
\label{vesicle}
\vec{R}(\theta,\varphi,t) = R(\theta,\varphi,t)\vec{e}_{r}
                     = R_{0} [\;1 + u(\theta,\varphi,t)\;]\vec{e}_{r} \;,  
\end{equation}
where $R_{0}$ is the radius of the spherical equilibrium shape, and 
$u(\theta,\varphi,t)$ describes an arbitrary, small shape perturbation.  The two
local tangent vectors on the quasi-spherical surface are then given by
\begin{eqnarray}
&& \vec{t}_{\theta} \equiv \D{\vec{R}}{\theta} = R \vec{e}_{\theta} + R_{\theta}\vec{e}_{r}
\nonumber \\
&& \vec{t}_{\varphi} \equiv \D{\vec{R}}{\varphi} = R\sin\theta \vec{e}_{\varphi} +
R_{\varphi}\vec{e}_{r} 
\;,
\end{eqnarray}
where $R_{\theta} \equiv \partial R /\partial \theta = R_0\; \partial u /\partial \theta$ and 
$R_{\varphi} \equiv \partial R /\partial \varphi = R_0\; \partial u /\partial \varphi$.  

It follows from Eq.(\ref{decomposition}) that for this geometry, the covariant
components of the velocity of each monolayer can be found as
\[ 
U^{\alpha}_{\pm} = W^{\alpha}_{\pm} 
           + a^{\alpha \beta} (\vec{t}_{\beta}\cdot \vec{e}_{r}) \partial_{t}R 
           = W^{\alpha}_{\pm} + a^{\alpha \beta} R_{\beta} \partial_{t}R \;. 
\]
Clearly, in the scheme of linearization where quantities are expressed accurate only to
first order in the perturbation represented by $u$, 
\begin{equation}
\label{components}
\vec{U}^{\pm}_{\mathrm t} = U^{\alpha}_{\pm} \vec{t}_{\alpha}
= R_{0}\;W^{\theta}_{\pm} \vec{e}_{\theta} 
+ (R_{0}\sin\theta)\;W^{\varphi}_{\pm} \vec{e}_{\varphi}\;.
\end{equation}
The components of $\vec{U}_{\pm} = U_{\theta}^{\pm}\vec{e}_{\theta}
+ U_{\varphi}^{\pm}\vec{e}_{\varphi} + U_{r}^{\pm}\vec{e}_{r}$
will also be needed later and are expressed here as well to first order in the
perturbation
\begin{eqnarray}
\label{physicalcomponents}
&& U_{\theta}^{\pm}(\theta,\varphi,t) = R_0 W^{\theta}_{\pm}(\theta,\varphi,t) \;, \nonumber \\
&& U_{\phi}^{\pm}(\theta,\varphi,t) = R_0 \sin\theta W^{\varphi}_{\pm}(\theta,\varphi,t) \;, 
\nonumber \\
&& U_{r}^{\pm}(\theta,\varphi,t) = R_0 \dot{u}(\theta,\varphi,t) \;,
\end{eqnarray}
where we have introduced a short-hand notation, $\dot{g}(t) = \partial g(t)/\partial t$,
for the time derivative of any function $g(t)$.
Another quantity to collect here is $D_{\alpha}W^{\alpha}_{\pm}$.  Within the
scheme of linearization,
\begin{equation}
\label{divergence}
D_{\alpha}W^{\alpha}_{\pm} = D_{\alpha}^{(0)}W^{\alpha}_{\pm} 
= \left( \D{W_\pm^\theta}{\theta} + \cot \theta W_\pm^\theta 
       + \D{W_{\pm}^{\varphi}}{\varphi} \right)  \;.
\end{equation}

Two differential operators will be defined here for later use, 
\begin{equation}
\vec{\nabla}_{\mathrm{L}} \equiv \vec{e}_{\theta} \frac{\partial}{\partial \theta}
+ \vec{e}_{\varphi} \frac{1}{\sin\theta}\cdot\frac{\partial}{\partial \varphi} \;,
\end{equation} 
and $\hat{L}^{2} \equiv -\vec{\nabla}_{\mathrm{L}} \cdot \vec{\nabla}_{\mathrm{L}}$.
Clearly, by invoking $\hat{L}^{2}$ on the spherical harmonics, $\mathcal{Y}_{\lmode m}$,
we have
\begin{equation}
\hat{L}^{2} \mathcal{Y}_{\lmode m} = \lmode (\lmode+1) \mathcal{Y}_{\lmode m} \;.
\end{equation}

Finally, we write down the expansions of all the relevant fields, $u(\theta,\varphi,t)$
and $\phi^{\pm}(\theta,\varphi,t)$ in the basis of the spherical harmonics,
$\mathcal{Y}_{\lmode m}(\theta,\varphi)$, 
\begin{eqnarray}
u(\theta,\varphi,t) &=& {\sum_{\lmode,m}}^{\prime} u_{\lmode m}(t) \; \mathcal{Y}_{\lmode m}\nonumber\\
& \equiv &\sum_{\lmode=2}\sum_{m=-\lmode}^{m=\lmode} u_{\lmode m}(t) \;\mathcal{Y}_{\lmode m}  \;,
\label{u_expansion}\\
\phi^{\Delta}(\theta,\varphi,t) &=& \phi^{\Delta}_{0} + 
{\sum_{\lmode,m}}^{\prime} \psi^{\Delta}_{\lmode m}(t) \; \mathcal{Y}_{\lmode m}\nonumber\\
& \equiv &\phi^{\Delta}_{0} 
+ \sum_{\lmode=2}\sum_{m=-\lmode}^{m=\lmode} \psi^{\Delta}_{\lmode m}(t) \; \mathcal{Y}_{\lmode m}
 \;, 
\label{phi_expansion1}\\
\phi^{\Sigma}(\theta,\varphi,t) &=& \phi^{\Sigma}_{0} + 
{\sum_{\lmode,m}}^{\prime} \psi^{\Sigma}_{\lmode m}(t) \; \mathcal{Y}_{\lmode m}\nonumber\\
& \equiv &\phi^{\Sigma}_{0} + \sum_{\lmode=2}\sum_{m=-\lmode}^{m=\lmode} \psi^{\Sigma}_{\lmode m}(t) \; 
\mathcal{Y}_{\lmode m} \;.  
\label{phi_expansion2}                      
\end{eqnarray}
where two alternative fields
\begin{eqnarray}
\phi^{\Delta}(\theta,\varphi,t) \equiv
\frac{\phi^{+}(\theta,\varphi,t) - \phi^{-}(\theta,\varphi,t)}{2} \;,\\
\phi^{\Sigma}(\theta,\varphi,t) \equiv
\frac{\phi^{+}(\theta,\varphi,t) + \phi^{-}(\theta,\varphi,t)}{2} \;,
\end{eqnarray}
are used instead of $\phi^{\pm}(\theta,\varphi,t)$ for later convenience.  
$\phi^{\Delta}_{0} \equiv (\phi^{+}_{0} - \phi^{-}_{0})/2$, and 
$\phi^{\Sigma}_{0} \equiv (\phi^{+}_{0} + \phi^{-}_{0})/2 = 0$ are constants
characterizing the equilibrium state of the vesicle; and
$\{u_{\lmode m}(t),\psi^{\Delta}_{\lmode m}(t),\psi^{\Sigma}_{\lmode m}(t) \}$ represent
the perturbations.

The reasons for not including $\lmode=0,1$ modes in the above expansions may not all
appear obvious.  The $\lmode =0$ mode is excluded in Eq.(\ref{u_expansion}) because
we will not be considering this mode. This is related to the fact that the dynamics
of this mode in the fluctuation spectrum is not only slaved by the dynamics of the
other modes, but is also unresolvable experimentally.  The $\lmode = 1$ mode in the
perturbation expansion corresponds to a simple translation of the vesicle and its amplitude
is thus set to zero without loss of generality.  The exclusion of the $\lmode = 1$ modes
in Eq.(\ref{phi_expansion1}) and Eq.(\ref{phi_expansion2}) is due to the fact that they
are not coupled to the shape changes of the vesicle within the scheme of small
perturbations.  

\subsection{{\sffamily Approximations}}
In the application that will be presented in the following, the most significant
approximation which has been made is that the near-equilibrium dynamics of
the vesicles are purely dissipative, or overdamped.  In other words, the inertial
effects in both the bulk hydrodynamics and the monolayer surface hydrodynamics are
neglected. This approximation has apparently been made in all theoretical work on
vesicle dynamics, based on the estimates put forward first by Milner and Safran~\cite{milner87}.
Underlying the consideration of Milner and Safran was, however, the assumption that
the only relevant dynamics of a membrane vesicle was the dynamics of the membrane
conformation. Therefore, although we adopt the same approximation in this paper,
we feel that additional considerations to those that led to the earlier estimates
are needed, given the fact that in addition to the surface geometry field monolayer
density fields are also being considered.  Thus, we briefly state this issue here.

Let's consider the bulk hydrodynamics first.  Making the approximation amounts
to neglecting both the inertial term and the non-linear convective term in the
Navier-Stokes equation written in Eq.(\ref{bulk}).  By estimating the respective
ratios of the two terms to the viscous-force term, we may have a guideline on
whether the approximation is justified or not.  The ratio of the inertial term
to the viscous-force term can be measured roughly by a dimensionless quantity
\begin{equation}
\label{Reynolds}
\bar{\mathcal{R}} \equiv \frac{\rho_{\rm b}L^{2}}{\eta t_{0}} 
                \sim \frac{\rho_{\rm b} |\partial\vec{v}/\partial t|}
                      {\eta |\nabla^{2} \vec{v}|}  \;,
\end{equation}
where $L$ and $t_{0}$ are, respectively, the typical length and time scales
characterizing the spatial and temporal variations of $\vec{v}$.  Another
dimensionless quantity measures the ratio of the convective term to the
viscous-force term,
\[
\mathcal{R} \equiv \frac{\rho_{\rm b}vL}{\eta} 
        \sim \frac{\rho_{\rm b} |(\vec{v} \cdot \nabla)\vec{v}|}
             {\eta |\nabla^{2} \vec{v}|} \;.
             \]
This quantity is the conventionally defined Reynolds number~\cite{landau}.

In the case of vesicle dynamics, motions in the bulk fluids are induced by motions
of the vesicle membranes, which may be considered as a linear composition of modes
of motion characterized by different wavelengths.  For each given mode
$L=\alpha \; \mu{\mathrm m}$ should correspond to the characteristic wavelength of
the mode.  For a vesicle of $20 \mu{\mathrm m}$ radius, $\alpha$ may range from
1 to 60, covering those modes that can be resolved under optical microscope.  For
a self-consistency check of the no-inertia approximation, the time
scale $t_{0}$ should represent the shortest time scale characterizing the overdamped
dynamics of a vesicle, rather than simply the time scale associated with the
overdamped relaxation of a pure bending mode, as was the case considered by Milner
and Safran~\cite{milner87}.  For pedagogical reasons, we will leave the quantitative
self-consistency check to the discussion, where more understanding of the various
relaxation time scales will have become available. 

To estimate the order of magnitude of $\mathcal{R}$, we replace $v$ by $l/t_{0}$,
where $l$ should be a measure of the shape deformations of the vesicles from their
equilibrium shapes, and is always much smaller than $L$ in the cases that we are
interested in.  Consequently, we have $\mathcal{R}/\bar{\mathcal{R}} = l/L \ll 1$.
Based on this analysis, we may always neglect the convective term in our considerations
of near-equilibrium dynamics of vesicles, provided that $\mathcal{R}$ is smaller
than 1.  Equations (\ref{bulk}) thus become
\begin{eqnarray}
\label{bulkequations}
&& \nabla \cdot \vec{v}^{\mathrm{(a)}} = 0 \;, \nonumber \\
&& \nabla p^{\mathrm{(a)}} = \eta \nabla^{2} \vec{v}^{\mathrm{(a)}} \;,
\end{eqnarray}

Similar order-of-magnitude analysis can be made regarding the approximation of 
neglecting the inertial and convective terms in the two-dimensional counterpart
of the bulk Navier-Stokes equation, Eq.(\ref{surfacedynamics}).  The order of 
magnitude of the surface inertial term is given by $\rho_{\mathrm m}|\vec{U}|/t_{0}$,
where $\rho_{\mathrm m}$ is the surface mass density of a lipid monolayer, 
and the order of magnitude of the forces acting on a monolayer may be represented
by $|\vec{T}| \sim \eta v/L \sim \eta |\vec{U}|/L$.  The ratio of these two terms
is then given by 
\begin{equation}
\mathcal{R}_{\mathrm m} = \frac{\rho_{\mathrm m}L}{\eta t_{0}}
                  = \frac{\rho_{\mathrm m}/L}{\rho_{\rm b}} \bar{\mathcal{R}} \;. 
\end{equation}
It is easy to work out that $\mathcal{R}_{\mathrm m} \ll \bar{\mathcal{R}}$.  Thus,
The approximation that the monolayer surface hydrodynamics is overdamped is well justified.
Under this approximation, Eq.(\ref{surfacedynamics}) reduces to 
\begin{equation}
\label{forcebalance}
\vec{f}^{\pm} = \vec{f}_{\mathrm{rs}}^{\pm} + \vec{T}^{\pm} 
             \mp b \; ( \vec{U}^{+}_{\mathrm{t}} - \vec{U}^{-}_{\mathrm{t}} ) = 0 \;.
\end{equation} 

It is more useful to decompose the above vector equations into their components
in the normal and the tangential directions.  Moreover, it turns out that, in the
normal direction, only the sum of the two monolayer equations is relevant, which
reads 
\begin{equation}
\label{normalforce}
( \vec{f}_{\mathrm{rs}}^{+} + \vec{f}_{\mathrm{rs}}^{-} ) \cdot \vec{n}
+ ( \vec{T}^{+} + \vec{T}^{-} ) \cdot \vec{n} = 0  \;.
\end{equation}
In the tangential directions, the more convenient expressions can be obtained
by the difference and the sum of the two monolayer equations, which look like
\begin{equation}
 ( \vec{f}_{\mathrm{rs,t}}^{+} - \vec{f}_{\mathrm{rs,t}}^{-} )
+ ( \vec{T}_{\mathrm t}^{+} - \vec{T}_{\mathrm t}^{-} ) 
- 2b\; ( \vec{U}^{+}_{\mathrm{t}} - \vec{U}^{-}_{\mathrm{t}} ) = 0 \;, \label{tangential1}
\end{equation}
\begin{equation}
 ( \vec{f}_{\mathrm{rs,t}}^{+} + \vec{f}_{\mathrm{rs,t}}^{-} )
+ ( \vec{T}_{\mathrm t}^{+} + \vec{T}_{\mathrm t}^{-} ) = 0 \;. \label{tangential2}
\end{equation}

In principle, the last three equations are applicable to vesicles with arbitrary
shapes.  In applying these equations to the case of a quasi-spherical
vesicle, we will only consider contributions which are first order in the 
perturbations.  The first-order contributions from both $\vec{f}_{\mathrm{rs}}^{\pm}$
and $\vec{U}^{\pm}_{\mathrm{t}}$ can be obtained exactly without approximation.  The
situation concerning $\vec{T}^{\pm}$ is not as well controlled, since an exact
solution would involve solving the bulk hydrodynamic equations for boundaries
of arbitrary shapes.  A practical approach, which is consistent with the
linearization scheme, is to make an approximation,
\begin{equation}
\label{viscousforce}
\vec{T}^{\pm} = \vec{T}^{\pm}_{0} ;,
\end{equation}
where $\vec{T}^{\pm}_{0}$ is obtained by solving the bulk hydrodynamic equations
given in Eq.(\ref{bulkequations})
for spherical boundaries that coincide with the spherical equilibrium shape of 
the vesicle.  

\subsection{{\sffamily Mechanical restoring forces}}
The general expressions for the mechanical restoring forces given in 
Eq.(\ref{restoringforce}) should be evaluated now for the specific case
of the quasi-spherical vesicle described by Eq.(\ref{vesicle}) to first order
in the perturbation fields.  

In the scheme of linearization, the mean curvature $H$ and the Gaussian curvature
$K$ take on the following forms,
\begin{eqnarray}
H&=&-\frac{1}{R_0}
-\frac{1}{2R_0} {\sum_{\lmode,m}}^{\prime} (\lmode+2)(\lmode-1) u_{\lmode m}
\mathcal{Y}{\lmode m} \;,\\
K&=&\frac{1}{R_0^2}
+\frac{1}{R_0^2} {\sum_{\lmode,m}}^{\prime} (\lmode+2)(\lmode-1) u_{\lmode m}
\mathcal{Y}{\lmode m} \;.
\end{eqnarray}
The Laplace-Beltrami operator $\Delta$ and the surface gradient operator 
$\vec{\nabla}_{\mathrm t}$ become
\[
\Delta = -\frac{\hat{L}^2}{R_0^2}\,,\quad
\vec{\nabla}_{\mathrm t}\Phi(\theta,\varphi) = 
\frac{1}{R_0}\vec{\nabla}_{\mathrm L}\Phi(\theta,\varphi) \;,
\]
if $\Phi(\theta,\varphi)$ is any function of first order in the perturbation fields.

By the use of the above simplified expressions, the sum of the two monolayer
normal components is evaluated to be 
\begin{align}
\begin{split}
\label{rsnormal}
&(\vec{f}_{\mathrm{rs}}^{+} + \vec{f}_{\mathrm{rs}}^{-} ) \cdot \vec{n}  \\
&= -\frac{2\bar{\tau}}{R_{0}}  
 - {\sum_{\lmode,m}}^{\prime} \Big\{
  \frac{\kappa_{\mathrm{eff}}}{R^3_0} E_{\lmode} u_{\lmode m}-\frac{4k_{\mathrm{eff}}}{R_0} \psi^\Sigma_{\lmode m} \\
&  -\left[ 4\frac{k_{\mathrm{eff}}}{R_0} \phi^{\Delta}_{0}
        +\frac{\lambda}{R_0^2} (\lmode+2)(\lmode-1) \right] \psi^\Delta_{\lmode m}                           \Big\} \mathcal{Y}_{\lmode m} \;,
\end{split}
\end{align}
where 
\begin{equation}
\bar{\tau} \equiv \sigma_{0} + \frac{\lambda\phi^{\Delta}_{0}}{R_0}
                    - k_{\mathrm{eff}} \left( \phi^{\Delta}_{0} \right)^2 
                     \; ,
\end{equation}
and 
\begin{equation}
\label{E_l}
E_{\lmode} \equiv (\lmode+2)(\lmode-1) \left[ \lmode (\lmode+1) 
+ \tau_{0} \; \frac{R_{0}^{2}}{\kappa_{\mathrm{eff}}} \right] \;,
\end{equation}
with 
\begin{equation}
\tau_{0} \equiv \bar{\tau} + \frac{\lambda\phi^{\Delta}_{0}}{R_0} \;.
\end{equation} 

The two tangential forces, $\vec{f}_{\mathrm{rs,t}}^{+}$ and 
$\vec{f}_{\mathrm{rs,t}}^{-}$, are, to first order in the perturbations, given by 
\[
\vec{f}_{\mathrm{rs,t}}^{\pm} =
- k_{\mathrm{eff}} (1+\phi_{0}^{\pm}) \vec{\nabla}_{\mathrm{L}} \phi^{\pm}
\mp \lambda (1 + \phi^{\pm}_{0}) \vec{\nabla}_{\mathrm{L}} H \;.
\]
The really relevant quantities are not the tangential forces themselves, but
the following ones,
\begin{align}
\begin{split}
\label{rstangent1}
\vec{\nabla}_{\mathrm{L}}\cdot ( \vec{f}_{\mathrm{rs,t}}^{+} - \vec{f}_{\mathrm{rs,t}}^{-} )
 =  {\sum_{\lmode,m}}^{\prime}  \lmode (\lmode+1) \left[
 2\frac{k_{\mathrm{eff}}}{R_0^2} \;
                              \psi^{\Delta}_{\lmode m} \right. \\
 \left. -\frac{\lambda}{R_0^3} 
                    (\lmode+2)(\lmode-1)u_{\lmode m}
+\frac{2k_{\mathrm{eff}}\phi^{\Delta}_{0}}{R^2_0} \psi^{\Sigma}_{\lmode m} \right]
\mathcal{Y}_{\lmode m} \; , 
\end{split}
\end{align}
and
\begin{align}
\begin{split}
\label{rstangent2}
\vec{\nabla}_{\mathrm{L}}\cdot ( \vec{f}_{\mathrm{rs,t}}^{+} + \vec{f}_{\mathrm{rs,t}}^{-} )
 =  {\sum_{\lmode,m}}^{\prime} \lmode (\lmode+1) \left[
 2\frac{k_{\mathrm{eff}}}{R_0^2} 
                               \psi^{\Sigma}_{\lmode m} \right. \\
 \left. -\frac{\lambda \phi^{\Delta}_{0}}{R_0^3} (\lmode+2)(\lmode-1) u_{\lmode m}                             
+\frac{2k_{\mathrm{eff}}\phi^{\Delta}_{0}}{R^2_0} \psi^{\Delta}_{\lmode m} \right]
\mathcal{Y}_{\lmode m} \; .
\end{split}
\end{align}

\subsection{{\sffamily Forces due to the bulk hydrodynamics}}
The forces, $\vec{T}^{\pm}_{0} $, defined in Eq.(\ref{viscousforce}) can be obtained
based on an adaptation to our problem of the classical Lamb solution \cite{lambsolution}.
This approach has already been used in the previous work on vesicle dynamics
by Schneider {\it et al.} \cite{schneider84}, Yeung and Evans \cite{evans95}, and
Seifert \cite{udo99}.  We will, therefore, only summarize the final results here. 

\subsubsection{{\sffamily The Lamb solution}}
In the Lamb solution, an alternative form of the boundary conditions,\linebreak
$\displaystyle \vec{v}^{\mathrm{(a)}}(r,\theta,\varphi,t)|_{r=R_{0}}= 
\vec{V}^{\mathrm{(a)}}(\theta,\varphi,t)$, is adopted, where the following three
scalar quantities are used instead of the three components of
$\vec{V}^{\mathrm{(a)}}(\theta,\varphi,t)$:
\begin{align}
\begin{split}
X^{\mathrm{(a)}}(\theta,\varphi,t) 
 & \equiv \vec{V}^{\mathrm{(a)}}(\theta,\varphi,t) \cdot \vec{e}_{r}\;,\\
Y^{\mathrm{(a)}}(\theta,\varphi,t)
& \equiv -R_0 \vec{\nabla} \cdot \vec{V}^{\mathrm{(a)}}(\theta,\varphi,t)\;,\\
Z^{\mathrm{(a)}}(\theta,\varphi,t)
& \equiv R_0 \vec{e}_{r} \cdot (\vec{\nabla} \times \vec{V}^{\mathrm{(a)}}(\theta,\varphi,t))
\;. 
\end{split}
\end{align}
The above three quantities can be expanded in the basis of the spherical harmonics,
\begin{eqnarray}
\displaystyle
X^{\mathrm{(a)}}(\theta,\varphi,t) 
& =& {\sum_{\lmode,m}}^{\prime} \; X^{\mathrm{(a)}}_{\lmode m}(t) \;\mathcal{Y}_{\lmode m} 
\;, \label{X}\\
Y^{\mathrm{(a)}}(\theta,\varphi,t)
& =&{\sum_{\lmode,m}}^\prime \; Y^{\mathrm{(a)}}_{\lmode m}(t) \;\mathcal{Y}_{\lmode m}
\;, \label{Y} \\
Z^{\mathrm{(a)}}(\theta,\varphi,t)
& = &{\sum_{\lmode,m}}^\prime \; Z^{\mathrm{(a)}}_{\lmode m}(t) \;\mathcal{Y}_{\lmode m}
\;. 
\end{eqnarray}
The normal and the tangential components of $\vec{T}^{\pm}_{0}$
can now be expressed in terms of $X^{\mathrm{(a)}}_{\lmode m}(t)$,
$Y^{\mathrm{(a)}}_{\lmode m}(t)$, and $Z^{\mathrm{(a)}}_{\lmode m}(t)$.  The normal
components read as 
\begin{eqnarray}
T_{0,\mathrm{n}}^{+} = 
-p_{0}^{\mathrm{(o)}} &+& \frac{\eta}{R_0} \sum_{\lmode=2} \sum_{m=-\lmode}^{\lmode}
\frac{-(2\lmode^{2}+3\lmode-2)}{\lmode+1}\; X^{\mathrm{(o)}}_{\lmode m}\;\mathcal{Y}_{\lmode m}\nonumber\\
    &+&\frac{\eta}{R_0} \sum_{\lmode=2} \sum_{m=-\lmode}^{\lmode}\frac{3}{\lmode+1}Y^{\mathrm{(o)}}_{\lmode m}\;\mathcal{Y}_{\lmode m} \;,
\nonumber\\
T_{0,\mathrm{n}}^{-} = 
p_{0}^{\mathrm{(i)}} &-& \frac{\eta}{R_0} \sum_{\lmode=2} \sum_{m=-\lmode}^{\lmode}
 \frac{2\lmode^2+\lmode-3}{\lmode}\;X^{\mathrm{(i)}}_{\lmode m} \; \mathcal{Y}_{\lmode m}\nonumber\\
     &-&\frac{\eta}{R_0} \sum_{\lmode=2} \sum_{m=-\lmode}^{\lmode}\frac{3}{\lmode}\;Y^{\mathrm{(i)}}_{\lmode m} \; \mathcal{Y}_{\lmode m} 
     \label{Tnormal}\;,
\end{eqnarray}
where $p_{0}^{\mathrm{(o)}}$ and $p_{0}^{\mathrm{(i)}}$ are the hydrostatic pressures
in the bulk fluids outside and inside the vesicle, respectively.

When the tangential components, $\vec{T}_{0,{\mathrm t}}^{\pm}$ are concerned, it is
more convenient to use $\vec{\nabla}_{\mathrm L} \cdot \vec{T}_{0,{\mathrm t}}^{\pm}$,
which look like
\begin{eqnarray}
&\vec{\nabla}_{\mathrm L} \cdot \vec{T}_{0,{\mathrm t}}^{+} =\nonumber\\
&\displaystyle{\frac{\eta}{R_0}}\sum_{\lmode=2} \sum_{m=-\lmode}^{\lmode} 
\left[ (\lmode+2) X^{\mathrm{(o)}}_{\lmode m} + (2\lmode+1)Y^{\mathrm{(o)}}_{\lmode m}
\right] \mathcal{Y}_{\lmode m} \;, \nonumber \\
&\vec{\nabla}_{\mathrm L} \cdot \vec{T}_{0,{\mathrm t}}^{-} =\\
&\displaystyle{\frac{\eta}{R_0}} \sum_{\lmode=2} \sum_{m=-\lmode}^{\lmode}
\left[ (\lmode-1)X^{\mathrm{(i)}}_{\lmode m} + (2\lmode+1) Y^{\mathrm{(i)}}_{\lmode m} 
\right] \mathcal{Y}_{\lmode m} \label{Ttangential} \;.\nonumber
\end{eqnarray}

\subsubsection{{\sffamily Surface continuity equations and kinematic matching}}
The ``amplitudes" $X^{\mathrm{(o)}}_{\lmode m}$, $X^{\mathrm{(i)}}_{\lmode m}$,
$Y^{\mathrm{(o)}}_{\lmode m}$ and $Y^{\mathrm{(i)}}_{\lmode m}$ can be directly
related to the amplitudes of the three perturbation fields, $u_{\lmode m}(t)$
and $\psi^{\pm}_{\lmode m}(t)$ as defined in Eqs.(\ref{u_expansion})-(\ref{phi_expansion2}).
The connections are provided by the two surface continuity equations given in
Eq.(\ref{surfacecontinuity}) and by the kinematic-matching conditions given in
Eq.(\ref{kinematic}).  Expressed to first order in the perturbations,
Eq.(\ref{surfacecontinuity}) read as
\begin{eqnarray}
D_{\alpha}W^\alpha_{\pm}& =& D_{\alpha}^{(0)}W^\alpha_{\pm}
= -\frac{\rho_0}{\rho^{\pm}_{0}}\dot{\phi}^\pm(\theta,\varphi,t)-2\dot{u}(\theta,\varphi,t)\nonumber\\
&=& - \frac{1}{1\pm \phi_{0}^{\Delta}}\; \dot{\phi}^\pm(\theta,\varphi,t)-2\dot{u}(\theta,\varphi,t) \; .\label{covariantW}
\end{eqnarray}
Substituting the kinematic-matching conditions Eq.(\ref{kinematic}) into Eq.(\ref{X}) and 
Eq.(\ref{Y}), using the expressions for $\vec{U}^{\pm}$ given in 
Eq.(\ref{physicalcomponents}), we arrive at
\begin{eqnarray}
 X^{\mathrm{(o)}}(\theta,\varphi,t) &=& X^{\mathrm{(i)}}(\theta,\varphi,t)
= R_{0}\; \dot{u}(\theta,\varphi,t) \;,\\
 Y^{\mathrm{(o)}}(\theta,\varphi,t) 
&=& -2 R_0\; \dot{u}(\theta,\varphi,t)\\ 
&&  - R_{0} \left[ \frac{1}{\sin\theta} \D{\,}{\theta} \left( W^{\theta}_{+} \cdot \sin\theta \right)
  + \D{W_{+}^{\varphi}}{\varphi} \right] \;, \nonumber\\
 Y^{\mathrm{(i)}}(\theta,\varphi,t) 
&=& -2 R_0\; \dot{u}(\theta,\varphi,t) \\
&&  - R_{0} \left[ \frac{1}{\sin\theta} \D{\,}{\theta} \left( W^{\theta}_{-} \cdot \sin\theta \right)
  + \D{W_{-}^{\varphi}}{\varphi} \right] \;. \nonumber 
\end{eqnarray}
Applying Eq.(\ref{covariantW}) to the last two equations, we finally have
\begin{eqnarray}
&& X^{\mathrm{(o)}}_{\lmode m}(t) = X^{\mathrm{(i)}}_{\lmode m}(t)
= R_{0}\;\dot{u}_{\lmode m}(t) \; , \nonumber \\
&& Y^{\mathrm{(o)}}_{\lmode m}(t) = \frac{R_{0}}{1+ \phi_{0}^{\Delta}}\; \dot{\psi}^{+}_{\lmode m}(t) \;,\nonumber\\
&&  Y^{\mathrm{(i)}}_{\lmode m}(t) =  \frac{R_{0}}{1- \phi_{0}^{\Delta}}\; \dot{\psi}^{-}_{\lmode m}(t) \;. 
  \label{matching}
\end{eqnarray}

By substituting Eq.(\ref{matching}) into Eq.(\ref{Tnormal}) and Eq.(\ref{Ttangential}),
we arrive at the following relevant expressions:
\begin{align}
\begin{split}
\label{Tnsum}
& T_{0,\mathrm{n}}^{+} + T_{0,\mathrm{n}}^{-} = 
 -(p_{0}^{\mathrm{(o)}}-p_{0}^{\mathrm{(i)}} ) \\
& + {\sum_{\lmode,m}}^{\prime} \frac{\eta}{\lmode (\lmode+1)} \bigg\{ 
  -(4\lmode^3+6\lmode^2-4\lmode-3)\dot{u}_{\lmode m} \\
&  -\frac{3[1+ (2\lmode +1)\phi_{0}^{\Delta}]}{1-(\phi_{0}^{\Delta})^{2}}
   \dot{\psi}^\Delta_{\lmode m} 
   +\frac{3(2\lmode+1+\phi_{0}^{\Delta})}{1-(\phi_{0}^{\Delta})^{2}}
   \dot{\psi}^\Sigma_{\lmode m} 
                      \bigg\} \mathcal{Y}_{\lmode m} \;, \\
\end{split}
\end{align}
\begin{eqnarray}
&&\vec{\nabla}_{\mathrm L} \cdot (\vec{T}_{0,{\mathrm t}}^{+}-\vec{T}_{0,{\mathrm t}}^{-})=
\frac{\eta}{R_{0}} {\sum_{\lmode,m}}^{\prime} \bigg[ 
3\dot{u}_{\lmode m}
+\frac{2(2\lmode+1)}{1-(\phi_{0}^{\Delta})^{2}} \dot{\psi}^{\Delta}_{\lmode m}\nonumber\\
&&-\frac{2(2\lmode+1)\phi_{0}^{\Delta}}{1-(\phi_{0}^{\Delta})^{2}} \dot{\psi}^{\Sigma}_{\lmode m}
 \bigg] 
\mathcal{Y}_{\lmode m} \;, \label{Tdifference} \\
&&\vec{\nabla}_{\mathrm L} \cdot (\vec{T}_{0,{\mathrm t}}^{+}+\vec{T}_{0,{\mathrm t}}^{-})=
\frac{\eta}{R_{0}} {\sum_{\lmode,m}}^{\prime} \bigg[ 
(2\lmode+1)\dot{u}_{\lmode m}\nonumber\\
&&-\frac{2(2\lmode+1)\phi_{0}^{\Delta}}{1-(\phi_{0}^{\Delta})^{2}} \dot{\psi}^{\Delta}_{\lmode m}
+\frac{2(2\lmode+1)}{1-(\phi_{0}^{\Delta})^{2}} \dot{\psi}^{\Sigma}_{\lmode m} \bigg] 
\mathcal{Y}_{\lmode m} \;. \label{Tsum}
\end{eqnarray}

\subsection{{\sffamily The intermonolayer frictional forces}}
The evaluation of the intermonolayer frictional forces appearing in 
Eq.(\ref{tangential1}) is straightforward. By applying $\vec{\nabla}_{\mathrm{L}}$
on $\vec{U}^{\pm}_{\mathrm t}$ given in Eq.(\ref{components}), we obtain, to first
order in the perturbations, 
\begin{eqnarray}
\vec{\nabla}_{\mathrm{L}} \cdot \vec{U}^{\pm}_{\mathrm t} &=& 
R_{0} \left( \D{W_{\pm}^{\theta}}{\theta} 
+ \cot\theta W_{\pm}^{\theta} + \D{W_{\pm}^{\varphi}}{\varphi} \right) \nonumber\\
&=&R_{0} \;(D_{\alpha} W_{\pm}^{\alpha} )=
-R_{0} \left[ \frac{\dot{\phi}^{\pm}}{1\pm \phi_{0}^{\Delta}} + 2\dot{u} \right] ,
\end{eqnarray}
where Eq.(\ref{divergence}) and Eq.(\ref{covariantW}) have been used.  It follows
immediately that
\begin{eqnarray}
\label{frictionfinal}
&&\vec{\nabla}_{\mathrm{L}} \cdot \left[ b (\vec{U}^{+}_{\mathrm t}-\vec{U}^{-}_{\mathrm t})
                            \right] 
= - bR_{0} (\frac{\dot{\phi}^{+}}{1+ \phi_{0}^{\Delta}} 
          -\frac{\dot{\phi}^{-}}{1- \phi_{0}^{\Delta}}) \\
&&= - 2 bR_{0} {\sum_{\lmode,m}}^{\prime} \; \left(
   \frac{1}{1-(\phi_{0}^{\Delta})^{2}} \dot{\psi}_{\lmode m}^{\Delta}
   - \frac{\phi_{0}^{\Delta}}{1-(\phi_{0}^{\Delta})^{2}} \dot{\psi}_{\lmode m}^{\Sigma}\; 
   \right) \mathcal{Y}_{\lmode m} \;.\nonumber
\end{eqnarray}

\subsection{{\sffamily Equations of motion }}
A set of 3 linear equations of motion can now be obtained for the three dynamic
variables characterizing a particular mode of perturbation, $u_{\lmode m}(t)$,
$\psi_{\lmode m}^{\Delta}(t)$ and $\psi_{\lmode m}^{\Sigma}(t)$.  Substitution
of both Eq.(\ref{rsnormal}) and Eq.(\ref{Tnsum}) into Eq.(\ref{normalforce}) yields
\footnote{The equation at the zeroth order is given by
$p_{0}^{\mathrm{(i)}} - p_{0}^{\mathrm{(o)}}
                 = 2\bar{\tau}/R_{0}$. }
\begin{align}            
\begin{split}
\label{bendingmode1}
&\frac{\eta}{\lmode (\lmode+1)} \bigg[ 
  -(4\lmode^3+6\lmode^2-4\lmode-3)\dot{u}_{\lmode m}\\
&  -\frac{3[1+ (2\lmode +1)\phi_{0}^{\Delta}]}{1-(\phi_{0}^{\Delta})^{2}} \dot{\psi}^\Delta_{\lmode m} 
  +\frac{3(2\lmode+1+\phi_{0}^{\Delta})}{1-(\phi_{0}^{\Delta})^{2}} \dot{\psi}^\Sigma_{\lmode m} 
                      \bigg]  \\
&= \frac{\kappa_{\mathrm{eff}}E_{\lmode}}{R^3_0} u_{\lmode m}
  -\left[ 4\frac{k_{\mathrm{eff}}}{R_0} \phi^{\Delta}_{0}
        +\frac{\lambda}{R_0^2} (\lmode+2)(\lmode-1) \right] \psi^\Delta_{\lmode m}\\
&  -\frac{4k_{\mathrm{eff}}}{R_0} \psi^\Sigma_{\lmode m} 
  \;.                 
\end{split}
\end{align}
Similarly, combining Eq.(\ref{rstangent1}), Eq.(\ref{Tdifference})
and Eq.(\ref{frictionfinal}) based on Eq.(\ref{tangential1}) leads to 
\begin{align}
\begin{split}
\label{differencemode1}
& \frac{\eta}{\lmode (\lmode+1)} \bigg\{ 
3\dot{u}_{\lmode m}+\left[2(2\lmode+1)+\frac{4bR_0}{\eta}\right]\\
&\cdot \left(
\frac{1}{1-(\phi_{0}^{\Delta})^{2}} \dot{\psi}_{\lmode m}^{\Delta}
   - \frac{\phi_{0}^{\Delta}}{1-(\phi_{0}^{\Delta})^{2}} \dot{\psi}_{\lmode m}^{\Sigma} \right)
                \bigg\} \\
& = \frac{\lambda}{R_0^2} (\lmode+2)(\lmode-1)u_{\lmode m}
    - 2\frac{k_{\mathrm{eff}}}{R_0} \psi^{\Delta}_{\lmode m} 
    - \frac{2k_{\mathrm{eff}}\phi^{\Delta}_{0}}{R_0} \psi^{\Sigma}_{\lmode m} 
\;,                                     
\end{split}
\end{align}
and combining Eq.(\ref{rstangent2}) and Eq.(\ref{Tsum}) according to Eq.(\ref{tangential2})
gives

\begin{align}
\begin{split}
\label{summode1}
& \frac{\eta}{\lmode (\lmode+1)} \bigg[ 
(2\lmode+1)\dot{u}_{\lmode m}
-\frac{2(2\lmode+1)\phi^{\Delta}_{0}}{1-(\phi_{0}^{\Delta})^{2}}\dot{\psi}^{\Delta}_{\lmode m}\\
&+\frac{2(2\lmode+1)}{1-(\phi_{0}^{\Delta})^{2}} \dot{\psi}^{\Sigma}_{\lmode m} \bigg] \\
&= 
  \frac{\lambda \phi^{\Delta}_{0}}{R_0^2} (\lmode+2)(\lmode-1) u_{\lmode m}
  - \frac{2k_{\mathrm{eff}}\phi^{\Delta}_{0}}{R_0} \psi^{\Delta}_{\lmode m}
  - 2\frac{k_{\mathrm{eff}}}{R_0} \psi^{\Sigma}_{\lmode m} \;.
\end{split}
\end{align}
In principle three independent modes of dynamics can now be determined by solving
the three equations of motion given above.  

A closer analysis of the equations shows, however, that only two modes of the three
are relevant on the time scales that are experimentally accessible, due to the fact
that there is an inherent separation of time scales involved in the problem.  Three
basic time scales may be defined based on Eq.(\ref{bendingmode1}),
Eq.(\ref{differencemode1}), and Eq.(\ref{summode1}), 
\begin{eqnarray}
&&t_{\mathrm c} = \frac{\eta R_{0}^{3}}{\kappa_{\mathrm{eff}}} \;
                \frac{\lmode^{3}}{\lmode (\lmode +1)E_{\lmode}}\;, \quad
t_{\Delta} = \frac{2bR_{0}^{2}}{k_{\mathrm{eff}}} \;
              \frac{1}{\lmode (\lmode +1)} \;,\nonumber\\
&&t_{\Sigma} = \frac{\eta R_{0}}{k_{\mathrm{eff}}} \;
              \frac{(2\lmode + 1)}{\lmode (\lmode +1)} \;.          
\end{eqnarray}
An order-of-magnitude estimate based on
\begin{eqnarray}\label{parameters}
&&\kappa_{\mathrm{eff}}=10^{-12}\;{\mathrm{erg}} \;,\quad
k_{\mathrm{eff}}= 30 \;\frac{\mathrm{erg}}{{\mathrm{cm}}^{2}}\;,\\
&&b= 5\times 10^{7}\frac{{\mathrm{erg}}\cdot{\mathrm{s}}} {{\mathrm{cm}}^{4}} \;,\quad R_{0}= 10 \; \mu {\mathrm m}\;,\quad \frac{\tau_{0} R_0^2}{\kappa_{\mathrm{eff}}}=10 \;,\nonumber
\end{eqnarray}
yields for $\lmode = 2$,
\[ t_{\mathrm c} \sim 0.2\; {\mathrm s} \;, \quad
   t_{\Delta} \sim 0.5 \;{\mathrm s}\;, \quad
   t_{\Sigma} \sim 2\times 10^{-7} \;{\mathrm s} \;.
\]
Clearly, the relaxation of $\psi^{\Sigma}_{\lmode m}$, which is characterized by
$ t_{\Sigma}$, is much faster than the relaxations of $u_{\lmode m}$ and 
$\psi^{\Delta}_{\lmode m}$.  Thus, we can assume 
\begin{equation}
\label{psi_Sigma}
\dot{\psi}^{\Sigma}_{\lmode m} \simeq 0 \;,  
\end{equation} 
over the time scales that characterize the relaxations of the slower modes.  Using
then Eq.(\ref{summode1}) to eliminate $\psi^{\Sigma}_{\lmode m}$ from the other two
equations, we arrive finally at two relevant equations of motion, written in the
following in a matrix form
\begin{equation}
\label{modeequation1}
\left( \begin{array}{c}\dot{u}_{\lmode m} \\
                  \dot{\psi}^\Delta_{\lmode m}
     \end{array} \right)
=\mathsf{A}_{2\times 2}^{-1} \mathsf{B}_{2\times 2}
 \left( \begin{array}{c} u_{\lmode m} \\
                    \psi^\Delta_{\lmode m} 
      \end{array}\right) 
\equiv - \mathsf{C}_{2\times 2}
 \left( \begin{array}{c} u_{\lmode m} \\
                    \psi^\Delta_{\lmode m} 
      \end{array}\right) \;.
\end{equation}
The two coefficient matrices are defined as
\begin{equation}
\mathsf{A}_{2\times 2} = 
\left( \begin{array}{ll} 
    \Gamma_{\lmode} & \quad \displaystyle{\frac{3-(2\lmode +1)\phi_{0}^{\Delta}}{1-(\phi_{0}^{\Delta})^{2}}} \\ 
    -3+ \beta_{\lmode} \phi^\Delta_0
     & \quad \displaystyle{-\frac{4b_0+2\beta_{\lmode}(1+(\phi_{0}^{\Delta})^{2})}{1-(\phi_{0}^{\Delta})^{2}}}
    \end{array} \right) \; ,
\end{equation}
and
\begin{equation}
\mathsf{B}_{2\times 2} = \frac{\lmode (\lmode+1)}{t_0}
\left( \begin{array}{ll}
-E_{\lmode} + 2\phi^{\Delta}_0 \lambda_{\lmode}  & \quad \lambda_{\lmode}  \\
 -\lambda_{\lmode} \left[ 1-
                    (\phi^\Delta_0)^{2} \right] &
                    \quad 
                    2k_0 \left[ 1-
                    (\phi^\Delta_0)^{2} \right]
     \end{array}
 \right)\ ,
\end{equation} 
where $t_{0} \equiv \eta R_{0}^{3}/\kappa_{\mathrm{eff}}$ defines a time scale, and
\[
b_{0} \equiv \frac{bR_{0}}{\eta} \;, \quad
\lambda_{0} \equiv \frac{\lambda R_{0}}{\kappa_{\mathrm{eff}}} \;, \quad
k_{0} \equiv \frac{k_{\mathrm{eff}}R_{0}^{2}}{\kappa_{\mathrm{eff}}} \;,
\]
are dimensionless parameters, and where a few short-hand notations have also been
defined 
\begin{eqnarray}
&&\Gamma_{\lmode} \equiv (2\lmode + 1) (2\lmode^{2} + 2\lmode -1) \;, \quad
 \beta_{\lmode} \equiv 2\lmode + 1 \;,\nonumber\\
&& \lambda_{\lmode} \equiv \lambda_{0} (\lmode -1) (\lmode +2) \;.
\end{eqnarray}

Further approximations can be made in the analysis of Eq.(\ref{modeequation1}).  The
dimensionless parameters $b_{0}$ and $k_{0}$ acquire rather large values for those
values of the physical parameters quoted in Eq.(\ref{parameters}), specifically,
$b_{0} \approx 5\times 10^{6}$, and $k_{0} \approx 3\times 10^{7}$, whereas 
parameter $\phi^\Delta_0$ is expected to be much smaller than one.  Thus, both
$\mathsf{A}_{2\times 2}$ and $\mathsf{B}_{2\times 2}$ can be simplified as
\begin{eqnarray}
\label{matrix_appr}
\mathsf{A}_{2\times 2} &\cong &
\left( \begin{array}{ll} 
    \Gamma_{\lmode} & \quad 0 \\ 
    0 & \quad -4b_0
    \end{array} \right)   \; , \nonumber \\
\mathsf{B}_{2\times 2} &\cong &
\frac{\lmode (\lmode+1)}{t_0}
\left( \begin{array}{ll}
-E_{\lmode} + 2\phi^{\Delta}_0 \lambda_{\lmode}  & \quad \lambda_{\lmode}  \\
 -\lambda_{\lmode}  & \quad 2k_0 
     \end{array}
 \right) \; .
\end{eqnarray} 
This yields a rather simple form for matrix $\mathsf{C}_{2\times 2}$,
\begin{eqnarray}
\mathsf{C}_{2\times 2} &\cong&
\frac{\lmode (\lmode+1)}{4b_{0}\Gamma_{\lmode}t_0}
\left( \begin{array}{ll}
4b_{0} \left( E_{\lmode} - 2\phi^{\Delta}_0 \lambda_{\lmode} \right)  & \quad -4b_{0} \lambda_{\lmode}  \\
 -\lambda_{\lmode} \Gamma_{\lmode} & \quad 2k_0 \Gamma_{\lmode} 
     \end{array}
 \right) \nonumber\\
&\equiv& 
\left( \begin{array}{ll}
\mathsf{C}_{11}  & \mathsf{C}_{12} \\
\mathsf{C}_{21} & \mathsf{C}_{22} 
     \end{array}  \right) \; .
\end{eqnarray} 

A comment is worth making here, which should help clarify the connections between
our theory and the previous theories on the vesicle dynamics \cite{milner87,schneider84,evans95}.
It concerns the two approximations made by use of Eq.(\ref{psi_Sigma}) and
Eq.(\ref{matrix_appr}).  The first approximation is actually equivalent to
imposing the so-called ``local area incompressibility constraint" that has been
used in some of the previous works \cite{milner87,schneider84,evans95}.  The physical
reason underlying the approximation, expressed in Eq.(\ref{psi_Sigma}), thus provides
a rationale for the use of the constraint.  However, it is incorrect, in our opinion,
to interpret in general, as the authors of Ref.~\cite{evans95} did, this constraint
as representing incompressibility of the monolayers as fluids, to be understood in
a way similar to the concept of incompressibility of bulk fluids, since the constraint
eliminates only one of the two mechanisms by which the monolayer density fields can change
-- the mechanism through changes of local surface geometry of the monolayers.  The second
approximation implies that, as far as the induced hydrodynamic motions in the bulk fluids
are concerned, the surface flow effects that are associated with the overall compressibility
of the monolayers are negligible.  This approximation has the same effect as the 
approximation used in Ref.~\cite{evans95} (the second equation in Eq.~(A.6) therein).    

\subsection{{\sffamily Dissipative dynamics of shape fluctuations}}
Finally, two dispersion relations, which characterize the two independent dissipative
modes of the vesicle dynamics, can be obtained as the two eigenvalues
of $\mathsf{C}_{2\times 2}$, 
\begin{eqnarray}
\label{dispersion} 
\Omega_{\pm}(\lmode) &=&
\frac{1}{2} \left[ \Omega_{\mathrm{c}}(\lmode) + \Omega_{\Delta}(\lmode) \right]\nonumber\\
&&\cdot\left[1 \pm
\sqrt{1-4\frac{\Omega_{\Delta}(\lmode) \cdot \Omega_{\mathrm{a}}(\lmode)}
            {\left[\Omega_{\mathrm{c}}(\lmode) 
            + \Omega_{\Delta}(\lmode)\right]^2}} \;
\right] \; ,
\end{eqnarray}
where

\begin{align}
\begin{split}
& \Omega_{\Delta}(\lmode) \equiv \mathsf{C}_{22} = 
\frac{\lmode (\lmode+1)k_{\mathrm{eff}}}{2 b R_0^2} \;, \\
&\Omega_{\mathrm{c}}(\lmode) \equiv \mathsf{C}_{11} =  \frac{\lmode (\lmode+1)(\lmode -1)(\lmode +2)}
                                 { \eta R_0^3 \Gamma_{\lmode}} 
                              \left[ \lmode (\lmode +1) \kappa_{\mathrm{eff}} 
                              + E_{\sigma} \right]  \;, \\
& \Omega_{\mathrm{a}}(\lmode) \equiv \frac{\lmode (\lmode+1)(\lmode -1)(\lmode +2)}
                                 { \eta R_0^3 \Gamma_{\lmode}} 
                              \left[ \lmode (\lmode +1) \kappa_{\mathrm{a}} 
                              + E_{\sigma} + \frac{\lambda^{2}}{k_{\mathrm{eff}}} \right]  \;.
\end{split}
\end{align}
In the above expressions, the definition
\begin{equation}
E_{\sigma} \equiv R_0^2 [\sigma_{0} - k_{\mathrm{eff}} (\phi^{\Delta}_{0})^{2} ]
\end{equation}
has been used, and    
\begin{equation}
\kappa_{\mathrm{a}} \equiv \kappa_{\mathrm{eff}} -
                       \frac{\lambda^{2}}{2k_{\mathrm{eff}}}
\end{equation}
defines a new bending rigidity, which will be called ``apparent" bending rigidity.

At this stage, the quantity of our main concern, namely, the time-dependent correlation function
of shape fluctuations of a quasi-spherical vesicle,
$\langle u_{\lmode m}^{*}(t) u_{\lmode m}(0) \rangle$, may be written generally as
\begin{equation}
\frac{\langle u_{\lmode m}^{*}(t) u_{\lmode m}(0) \rangle}{\langle |u_{\lmode m}|^{2} \rangle} 
= \beta_{-} e^{\displaystyle - \Omega_{-}t} + \beta_{+} e^{\displaystyle - \Omega_{+}t} \;.
\end{equation}
Clearly, $\beta_{-} + \beta_{+} = 1$.  To obtain analytical
expressions for the amplitudes, $\beta_{\mp}$, we employ Onsager's regression hypothesis that
the equations of motion governing the macroscopic near-equilibrium dynamics can be applied to
the dynamics of spontaneous fluctuations \cite{onsager}.  Following the formalism given in
Ref.~\cite{reichl} we thus have
\begin{equation}
\langle u_{\lmode m}^{*}(t) u_{\lmode m}(0) \rangle 
= \langle |u_{\lmode m}|^{2} \rangle \; (e^{\displaystyle -\mathsf{C}t})_{11}
 + \langle u_{\lmode m}^{*} \psi_{\lmode m}^{\Delta} \rangle \;
 (e^{\displaystyle -\mathsf{C}t})_{21} \; ,
\end{equation}
which leads to the following specific expressions for $\beta_{\mp}$:
\begin{eqnarray}
\beta_{-} & = & \frac{1}{\Omega_{+}-\Omega_{-}}
         \left[ \Omega_{\Delta}-\Omega_{-}
         - \mathsf{C}_{21} \frac{\langle u_{\lmode m}^{*} \psi_{\lmode m}^{\Delta} \rangle}
                           {\langle |u_{\lmode m}|^{2} \rangle} \right] \nonumber \\
      & = & \frac{1}{\Omega_{+}-\Omega_{-}}
         \left[ \Omega_{\Delta}-\Omega_{-}
         + \Omega_{\Delta} \; \left( \frac{\lambda (\lmode -1)(\lmode +2)}{2 k_{\mathrm{eff}}R_0}
                            \right)^{2}
         \right] \;, \nonumber \\  
\beta_{+} & = & \frac{1}{\Omega_{+}-\Omega_{-}}
         \left[ \Omega_{+}-\Omega_{\Delta}
         + \mathsf{C}_{21} \frac{\langle u_{\lmode m}^{*} \psi_{\lmode m}^{\Delta} \rangle}
                           {\langle |u_{\lmode m}|^{2} \rangle} \right] \\
       & = & \frac{1}{\Omega_{+}-\Omega_{-}}
         \left[ \Omega_{+}-\Omega_{\Delta}
         - \Omega_{\Delta} \; \left( \frac{\lambda (\lmode -1)(\lmode +2)}{2 k_{\mathrm{eff}}R_0}
                            \right)^{2}
                            \right] \;.\nonumber           
\end{eqnarray}
$\langle |u_{\lmode m}|^{2} \rangle$ and
$\langle u_{\lmode m}^{*} \psi_{\lmode m}^{\Delta} \rangle$ represent some of the
static correlation functions, the calculations of which are briefly sketched in
{\bf {\sffamily Appendix A}}.

\section{{\sffamily Discussion}}
The dispersion relations, Eq.(\ref{dispersion}), predicted by our theory appear
{\em formulistically} identical to those worked out by Yeung and Evans \cite{evans95}
and by Bivas {\it et al.} \cite{bivas99}.  But, our definitions of two of the three
quantities involved in the dispersion relations, namely, $\Omega_{\mathrm{c}}(\lmode)$
and $\Omega_{\mathrm{a}}(\lmode)$, differ in detail from those previous results.  
More importantly, we feel that interpretations of those physical parameters contained
in the dispersion relations, $\kappa_{\mathrm{eff}}$, $k_{\mathrm{eff}}$, $\lambda$ and
$E_{\sigma}$, need to be reconsidered.  

We have already remarked on the issue of interpretation in Section 2.2, when
introducing the effective free energy describing a vesicle with a coarse-grained
configuration characterized by $\vec{R}(u^{1},u^{2},t)$, and $\rho^{\pm}(u^{1},u^{2},t)$.
As we have not yet performed for systems of quasi-spherical vesicles the coarse-graining
procedure similar to those described in Refs.~\cite{cai95,peliti01}, we are only
able to give a semi-quantitative discussion of the effective mesoscopic parameters.

If we consider systems of quasi-spherical vesicles where fluctuations about
equilibrium shapes are small, where a Gaussian theory would suffice in describing
the small-scale fluctuations, we may argue that the renormalization of bending
rigidity due to nonlinearities \cite{rgpapers} can be neglected and that 
$\kappa_{\mathrm{eff}} \simeq \kappa_{\mathrm{m}} - \lambda_{\mathrm{m}}^2/2k_{\mathrm{m}}$,
where $\kappa_{\mathrm{m}}$, $\lambda_{\mathrm{m}}$, and $k_{\mathrm{m}}$ 
represent the phenomenological parameters corresponding the microscopic cut-off
length scale.  Note that the part subtracted from $\kappa_{\mathrm{m}}$ arises
from the coupling between membrane curvature field and the density-difference
field \cite{per98}.  But, making a quantitative statement about $\kappa_{\mathrm{eff}}$ is
not trivial, at least, not straightforward, even from the point of view of experiments.
In the so-called flicker-noise analysis experiments, a bending rigidity $\kappa$ and a
``surface-tension" constant $\Sigma_{\mathrm{e}}$ are inferred from fitting the
experimental data on $\langle |\hat{u}_{\lmode m}|^{2} \rangle$ onto a functional
expression, 
\begin{equation}
\label{experiment}
\langle |\hat{u}_{\lmode m}|^{2} \rangle = \frac{k_{\mathrm B}T}
       {(\lmode -1)(\lmode +2)\kappa [\lmode (\lmode + 1) + \Sigma_{\mathrm{e}} ]} \;.
\end{equation}
Based on our theory and calculations presented in Appendix A, which yields 
\begin{equation}
\label{static_u}
\langle |\hat{u}_{\lmode m}|^{2} \rangle = \frac{k_{\mathrm B}T}
       {(\lmode -1)(\lmode +2)\kappa_{\mathrm a} [\lmode (\lmode + 1) + \Sigma ]} \;,
\end{equation}
where 
\begin{equation}
\label{Sigma}
\Sigma \equiv \frac{1}{\kappa_{\mathrm{a}}}
      \left[ \sigma_{0}R_{0}^{2} - k_{\mathrm{eff}}R_{0}^{2} (\phi_{0}^{\Delta})^{2}
       + \frac{\lambda^{2}}{k_{\mathrm{eff}}} \right] \;,
\end{equation}
the experimentally obtained bending rigidity $\kappa$ should be identified not with
$\kappa_{\mathrm{eff}}$, but with $\kappa_{\mathrm{a}}$, the apparent or renormalized
bending rigidity. 

$k_{\mathrm{eff}}$ is expected to be lower than its microscopic counterpart due to
the renormalization effect when the vesicles are either in or close to the {\em floppy}
states, as we have already mentioned in Section 2.2.  It is our opinion that the values
of area compressibility moduli for various single-component membrane systems reported
so far in the literature \cite{rawicz,gawrisch} are actually the values of the microscopic
area compressibility moduli, as these values have been obtained when the membranes under
observation are in {\em stretched-tense} states \cite{peliti01}.  In the absence of any available
experimental data on $k_{\mathrm{eff}}$, we make an estimate based on the semi-quantitative
derivation of the parameter given in Ref.~\cite{peliti01} by using a short-distance
cut-off of optical lengths, $\Lambda^{-1} = 100 \;\mathrm{nm}$ and obtain
$k_{\mathrm{eff}}\simeq 30 \; \mathrm{erg}/\mathrm{cm}^{2}$.  This value is lower by 2-3 fold
than the values of the microscopic area compressibility moduli typically quoted in the
literature .

The parameter, $\lambda$, is a rather elusive one, due to the fact that understanding
of the physical origin of $\lambda_{\mathrm{m}}$ is limited.  We will follow the
discussion on it put forward by Seifert and Langer \cite{udo93} and put an estimate
on $\lambda_{\mathrm{m}}$ as $\lambda_{\mathrm{m}}=2k_{\mathrm{m}}d$, where
$d \simeq 1 \mathrm{nm}$, representing roughly half of the monolayer thickness.  Taking
$d \simeq 1 \; \mathrm{nm}$ and $k_{\mathrm{m}}=100\; \mathrm{erg}/\mathrm{cm}^{2}$,
we have $\lambda_{\mathrm{m}} \simeq 2 \times 10^{-5} \; \mathrm{erg}/\mathrm{cm}$.
It is not difficult to see, if one follows the kind of coarse-graining procedures
described in \cite{cai95}, that $\lambda \simeq \lambda_{\mathrm{m}}$.  

The parameter, $E_{\sigma}$, should according to our calculation be related to the
experimentally inferred parameter $\Sigma_{\mathrm{e}}$:
\begin{equation}
E_{\sigma} =\kappa_{\mathrm{a}} \Sigma_{\mathrm{e}} - \frac{\lambda^{2}}{k_{\mathrm{eff}}} 
             \;,  
\end{equation} 
as, based on Eq.(\ref{experiment}) and Eq.(\ref{static_u}), $\Sigma$ should be identified
with $\Sigma_{\mathrm{e}}$.  

Our analytical expressions for $\beta_{\mp}$ differ also qualitatively both from those
given in Ref.~\cite{evans95} and from those given in Ref.~\cite{bivas99}.  Explicitly, the
qualitative corrections to the expressions given in Ref.~\cite{evans95} are 
\[
\Delta \beta_{\mp} = \pm \frac{\langle |u_{\lmode m}|^{2} \rangle\; \Omega_{\Delta} }
                       {\Omega_{+}-\Omega_{-}}
                \; \left( \frac{\lambda (\lmode -1)(\lmode +2)}{2 k_{\mathrm{eff}}R_0}
                            \right)^{2}  \;.
\]
Quantitatively, however, the corrections turn out to be negligible, as a quick,
order-of-magnitude estimate by using the values of $\lambda$ and $k_{\mathrm{eff}}$
quoted above and $R_0 = 20 \;\mu \mathrm{m}$ shows.  But, the corrections to the
expressions obtained in Ref.~\cite{bivas99} become significant quantitatively, as we
will show numerically in the following.

Based on the above discussion, we now present a systematic numerical analysis of the
analytical results given in the previous section.  The analysis itself is obviously
straightforward, but it brings out some numerical consequences that are very relevant
to experimental studies of the dynamics of vesicle shape fluctuations.  We first
analyze two cases, where two important parameters take on different numerical values: 
case a) $k_{\mathrm{eff}} = 100 \; \mathrm{erg}/\mathrm{cm}^{2}$,
$b= 10^{7} \;\mathrm{erg}\cdot \mathrm{sec}/\mathrm{cm}^{4}$, and case b)
$k_{\mathrm{eff}} = 30 \; \mathrm{erg}/\mathrm{cm}^{2}$, and 
$b= 2\times 10^{8} \;\mathrm{erg}\cdot \mathrm{sec}/\mathrm{cm}^{4}$, whereas the other
parameters assume the same values in both cases (see the captions for the tables and figures).  
Table~1 together with Figure~1, and Table~2 together with Figure~2 illustrate the
numerical results, corresponding to case a) and case b), respectively.

\begin{table*}
\centering
\begin{tabular}{c||r|r|r|r|r|r|r|r|r}
\hline
$\lmode$ & 2 & 3 & 4 & 5 & 6 & 7 & 8 & 9 & 10 \\
\hline
$\tau_{-}[\mathrm{ms}]$ & 11525 & 4940 & 2648 & 1583 & 1018 & 691 & 490 & 359 & 271 \\
\hline
$\tau_{+}[\mathrm{ms}]$ & 132 & 66 & 39 & 26 & 19 & 14 & 11 & 9 & 7 \\
\hline 
$\beta_{-}$ & 0.994 & 0.987 & 0.981 & 0.976 & 0.970 & 0.965 & 0.960 & 0.954 & 0.949 \\
\hline
  \end{tabular}
  \caption{Numerical values of the time scales, $\tau_{\mp} \equiv 1/\Omega_{\mp}$, and
  the amplitude $\beta_{-}$, characterizing the two dissipative modes of vesicle dynamics
  for $\kappa_{\mathrm{a}}= 10^{-12}\; \mathrm{erg} = 24 k_{B}T$,
  $\eta=0.01\erg\s/\cm^3$,
  $\lambda = 2\times 10^{-5}\; \erg/\cm$, $\Sigma=10$, $R=20 \;\mu \m$,
  $k_{\mathrm{eff}}=100 \;\erg/\cm^2$, and $b=10^{7}\;\erg\s/\cm^4$.}
\label{table1}
\end{table*}

\begin{table*}
\centering
\begin{tabular}{c||r|r|r|r|r|r|r|r|r}
\hline
$\lmode$ & 2 & 3 & 4 & 5 & 6 & 7 & 8 & 9 & 10 \\
\hline
$\tau_{-}[\mathrm{ms}]$ & 31977 & 21796 & 15485 & 11378 & 8625 & 6720 & 5362 & 4366 & 3617 \\
\hline
$\tau_{+}[\mathrm{ms}]$ & 3185 & 995 & 448 & 241 & 146 & 95 & 65 & 47 & 35 \\
\hline
$\beta_{-}$ & 0.287 & 0.187 & 0.143 & 0.117 & 0.099 & 0.087 & 0.077 & 0.069 & 0.062 \\
\hline
  \end{tabular}
  \caption{Numerical values of the time scales, $\tau_{\mp} \equiv 1/\Omega_{\mp}$, and
  the amplitude $\beta_{-}$, characterizing the two dissipative modes of vesicle dynamics
  for $\kappa_{\mathrm{a}}= 10^{-12}\; \mathrm{erg} = 24 k_{B}T$,
  $\eta=0.01\erg\s/\cm^3$,
  $\lambda = 2\times 10^{-5}\; \erg/\cm$, $\Sigma=10$, $R_0=20 \;\mu \m$,
  $k_{\mathrm{eff}}=30 \;\erg/\cm^2$, and $b=2 \times 10^{8}\;\erg\s/\cm^4$. }
\label{table2}
\end{table*}

As the numerical results clearly show, case a) and case b) yield two rather different
scenarios for the relaxation dynamics of shape fluctuations of quasi-spherical vesicles.
In case a), where the values of $k_{\mathrm{eff}}$ and $b$ are chosen to conform to the
values quoted canonically in the literature, the time scales characterizing the two modes
are separated by almost two orders of magnitude, and the slower mode takes up almost all
of the full amplitude.  Given the typical temporal resolution of milliseconds and the typical
accuracy of 10\% in determining fluctuation amplitudes in standard flicker-noise analysis
experiments \cite{jonas}, only the slower mode can be effectively resolved by experimental
observations and analysis.  The different physical nature of the two modes are also illustrated
by Figure~1.  The dispersion relation describing the slower mode is well approximated
by $\Omega_{\mathrm{a}}(l)$, whereas that describing the faster mode is well approximated
by $\Omega_{\Delta}(l)$.  The functional form of $\Omega_{\mathrm{a}}(l)$ is the same as
that of the dispersion relation derived by Milner and Safran for the relaxation of a pure
bending mode in the absence of the coupling between membrane geometry and monolayer density
fields \cite{milner87}.  The consequence of the coupling is, however, present in our result.
It is indicated by the fact that the bending rigidity appearing in $\Omega_{\mathrm{a}}(l)$
is the apparent one, $\kappa_{\mathrm{a}}$, which is the result of renormalization of
$\kappa_{\mathrm{eff}}$ by fluctuations in the monolayer density fields in the presence of
the coupling.  The reason for the renormalization effect is also revealed in Fig.~1:
Fluctuations in the difference of the monolayer density fields relax much more quickly
than pure shape fluctuations, i.e. $\Omega_{\Delta}(l) \gg \Omega_{\mathrm{c}}(l)$.

\begin{figure}
\vspace*{.5cm}
\centerline{
\resizebox{9cm}{!}{
  \includegraphics{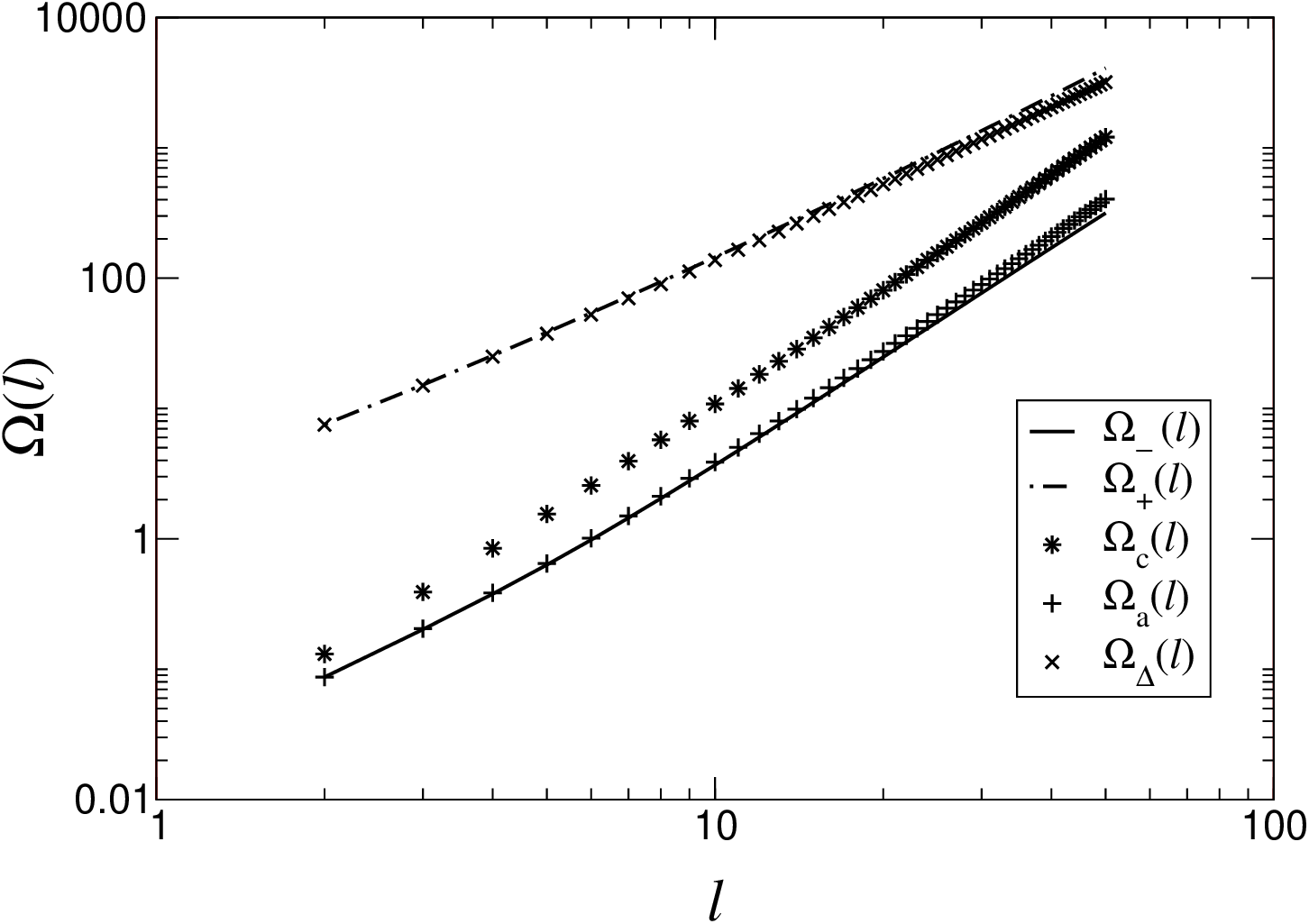}  
}}
\caption{Various dispersion relations relevant for describing the two dissipative modes of
vesicle dynamics
for $\kappa_{\mathrm{a}}= 10^{-12}\; \mathrm{erg} = 24 k_{B}T$,
  $\eta=0.01\erg\s/\cm^3$,
  $\lambda = 2\times 10^{-5}\; \erg/\cm$, $\Sigma=10$, $R=20 \;\mu \m$,
  $k_{\mathrm{eff}}=100 \;\erg/\cm^2$, and $b=10^{7}\;\erg\s/\cm^4$. }
\end{figure}

In case b), the quantitative differences between the two relaxation time scales are smaller
than in case a), and the slower mode takes up a much smaller fraction of the full amplitude.
Relevant to experimental studies is that, for the first few low $\lmode$ values, both relaxation
modes can be resolved with the typical experimental resolutions.  Another point to note is
that there is a reversal in the physical nature of the two modes in comparison with case a),
as illustrated by Figure~2.  It is now the faster mode that describes approximately the
relaxation of bending deformation, where the governing bending rigidity is, however,
$\kappa_{\mathrm{eff}}$ instead of $\kappa_{\mathrm{a}}$.  In other words, the renormalization
effect is absent.  This absence can also easily be rationalized: The relaxations of
fluctuations in the difference of the monolayer density fields require longer time
scales than the relaxations of pure shape fluctuations.  The slower mode no longer describes
relaxation of bending fluctuations, although it can not be approximated simply by
$\Omega_{\Delta}(l)$ either.  Clearly, case b) provides a new scenario that is different from
the ``conventional" scenario depicted in case a).  Indeed, a recent analysis of data obtained
from flicker-noise measurements of giant quasi-spherical vesicles made of
stearoyl-oleoyl-phosphatidylcholine \cite{pott}, which seemed to have resolved two modes with
corresponding time scales quantitatively similar to those quoted in Table~2, suggests reasons
for searching for this scenario.   

\begin{figure}
\vspace*{.5cm}
\centerline{
\resizebox{9cm}{!}{
  \includegraphics{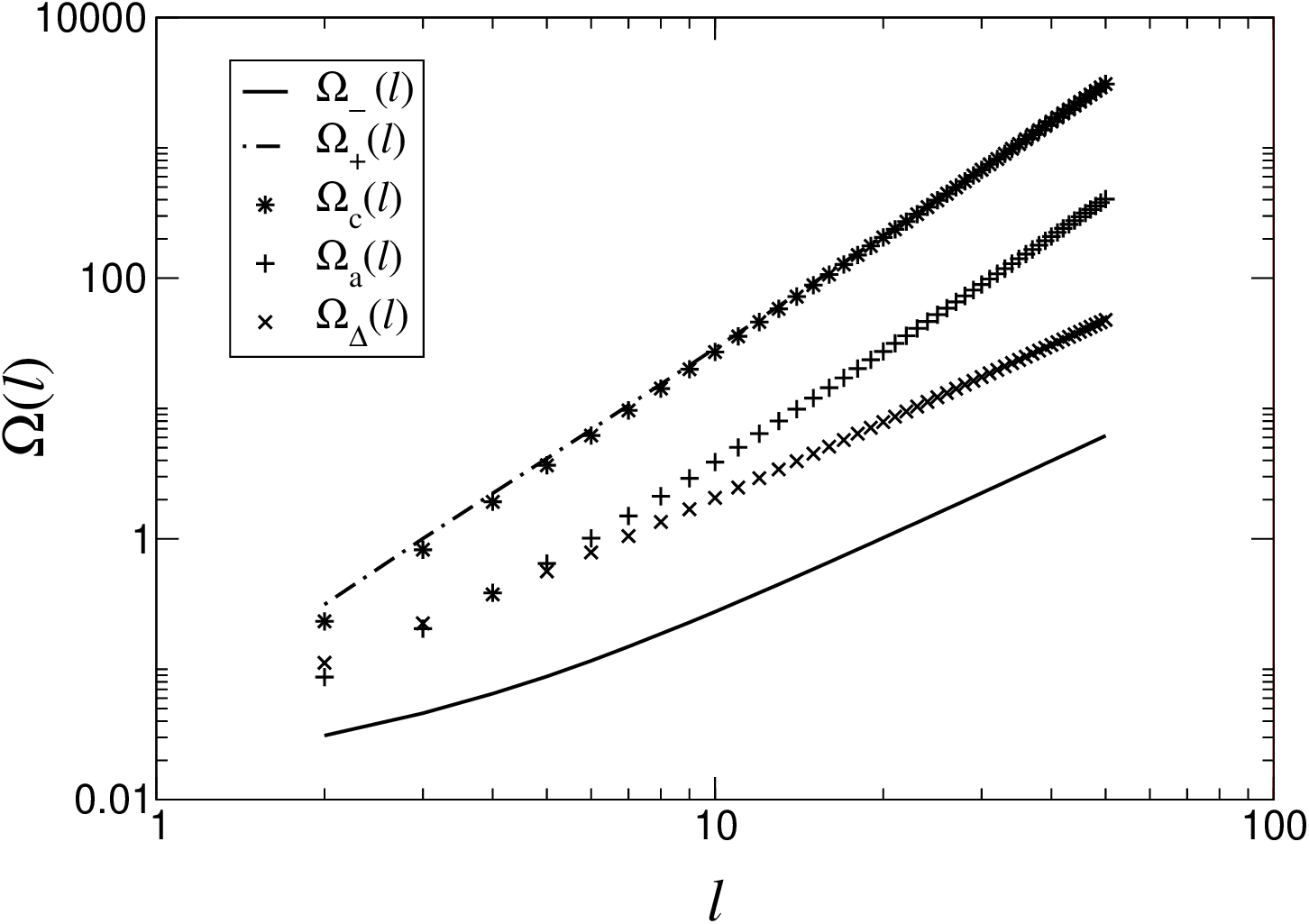}  
}}
\caption{The same dispersion relations as those illustrated in Fig.~1, for the same parameter
      values but the following two:
      $k_{\mathrm{eff}}=30 \;\erg/\cm^2$, and $b=2 \times 10^{8}\;\erg\s/\cm^4$.}
\end{figure}

We have in the above discussed our interpretation of the monolayer compressibility
modulus, $k_{\mathrm{eff}}$, and proposed that the values of this parameter should
be lower than the standard values quoted in the literature.  The numerical analysis
illustrates clearly possible consequences that a reduction in the value of
$k_{\mathrm{eff}}$ can have.  As a case of comparison with case b) (Fig.~2) Figure~3
displays the various dispersion relations obtained when the value of $k_{\mathrm{eff}}$
is increased to $100 \; \erg/\cm^{2}$ and all the rest of the parameters have the same
values as in case b).  Quantitatively, the increase in the value of $k_{\mathrm{eff}}$
reduces the characteristic time scales of both of the modes, affecting the slower mode
more significantly, though.  There is also a qualitative change in the nature of the
two modes for the first few low $\lmode$ values.   For the higher value of $k_{\mathrm{eff}}$,
the slower mode reflects predominantly the relaxation of bending deformation governed by
the renormalized bending rigidity $\kappa_{\mathrm{a}}$, and the faster mode reflects largely
the relaxation of the monolayer density-difference field, in reversal to case b).  This
point serves to underline the need to resolve quantitatively the issue of whether the
renormalization by fluctuations of compressibility moduli of lipid-bilayer membranes
in systems of giant vesicles is significant enough to be experimentally relevant.  
To be sure, reductions in the value of $k_{\mathrm{eff}}$ do not always cause the
kind of quantitative as well qualitative changes in the relaxation dynamics just
discussed.  In cases where the intermonolayer friction coefficient, $b$, has lower
values, e.g., $b = 10 ^{7} \; \erg \s/\cm^{4}$, the changes in the relaxation dynamics
of the two modes for low $\lmode$ values are not significant for the same increase
in the value of $k_{\mathrm{eff}}$.

\begin{figure}
\vspace*{.5cm}
\centerline{
\resizebox{9cm}{!}{
  \includegraphics{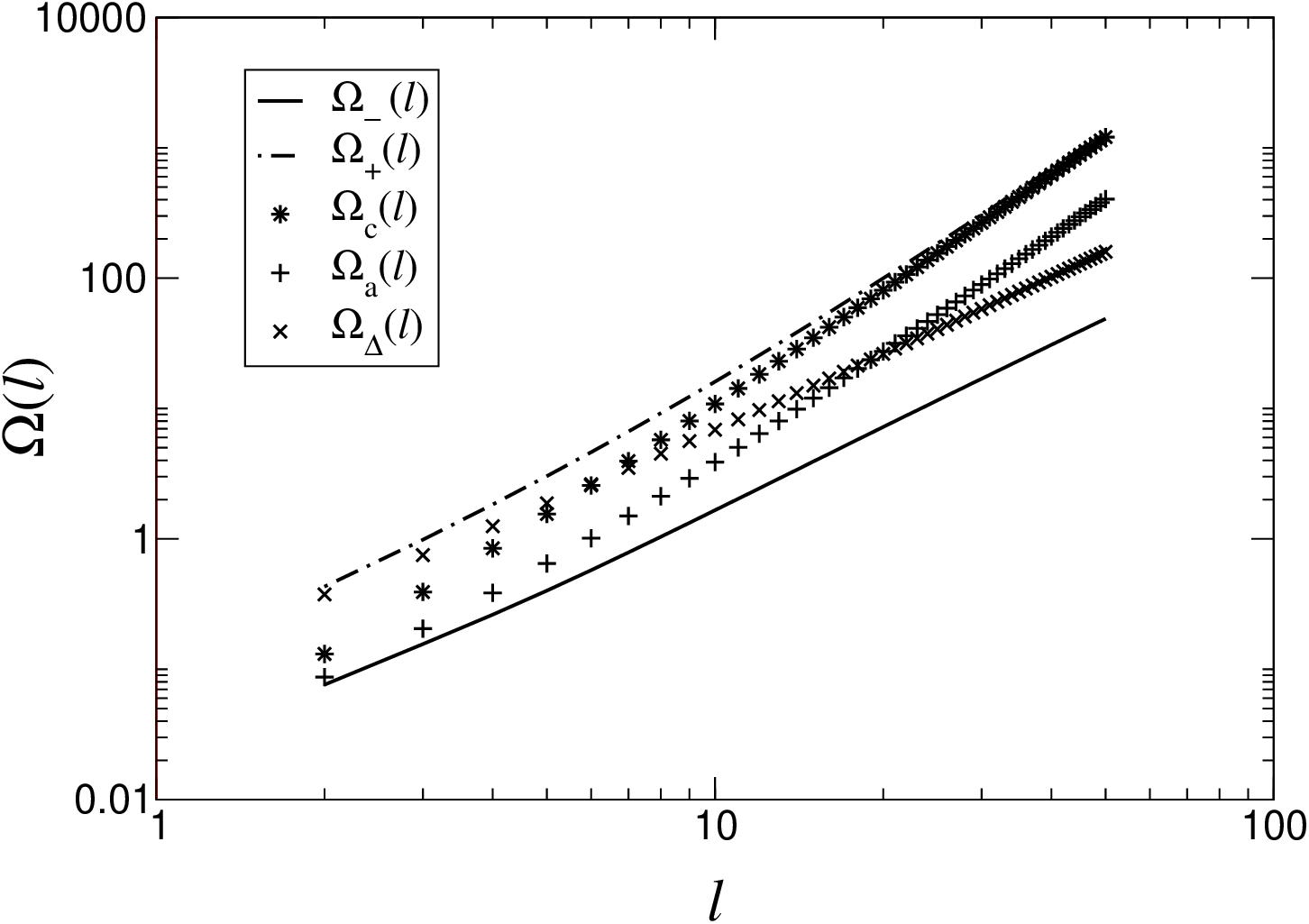}  
}}
\caption{The same dispersion relations as those illustrated in Fig.~2, for the same
parameter values but one: $k_{\mathrm{eff}}= 100 \; \erg/\cm^{2}$. }
\end{figure}

Another point to discuss in connection with experimental studies of vesicle fluctuations
by the technique of flicker-noise measurement and analysis concerns the effects that
variations in the parameter $\Sigma$ have on the relaxation dynamics of the fluctuations,
as $\Sigma$ is a vesicle-specific parameter that typically varies in the range from 
0 to 25 \cite{jonas}.  We have checked systematically in the numerical analysis
the effects, which are summarized by a comparison of Figure~4 with Figure~2.  It is
easy to see that a reduction in $\Sigma$ from 10 to 0 leads to a significant increase
in the characteristic time scale of the slower mode, but very small increase in the
time scale characterizing the faster mode.  Moreover, the relative positions of the
different dispersion relations do not change qualitatively.  

\begin{figure}
\vspace*{.5cm}
\centerline{
\resizebox{9cm}{!}{
  \includegraphics{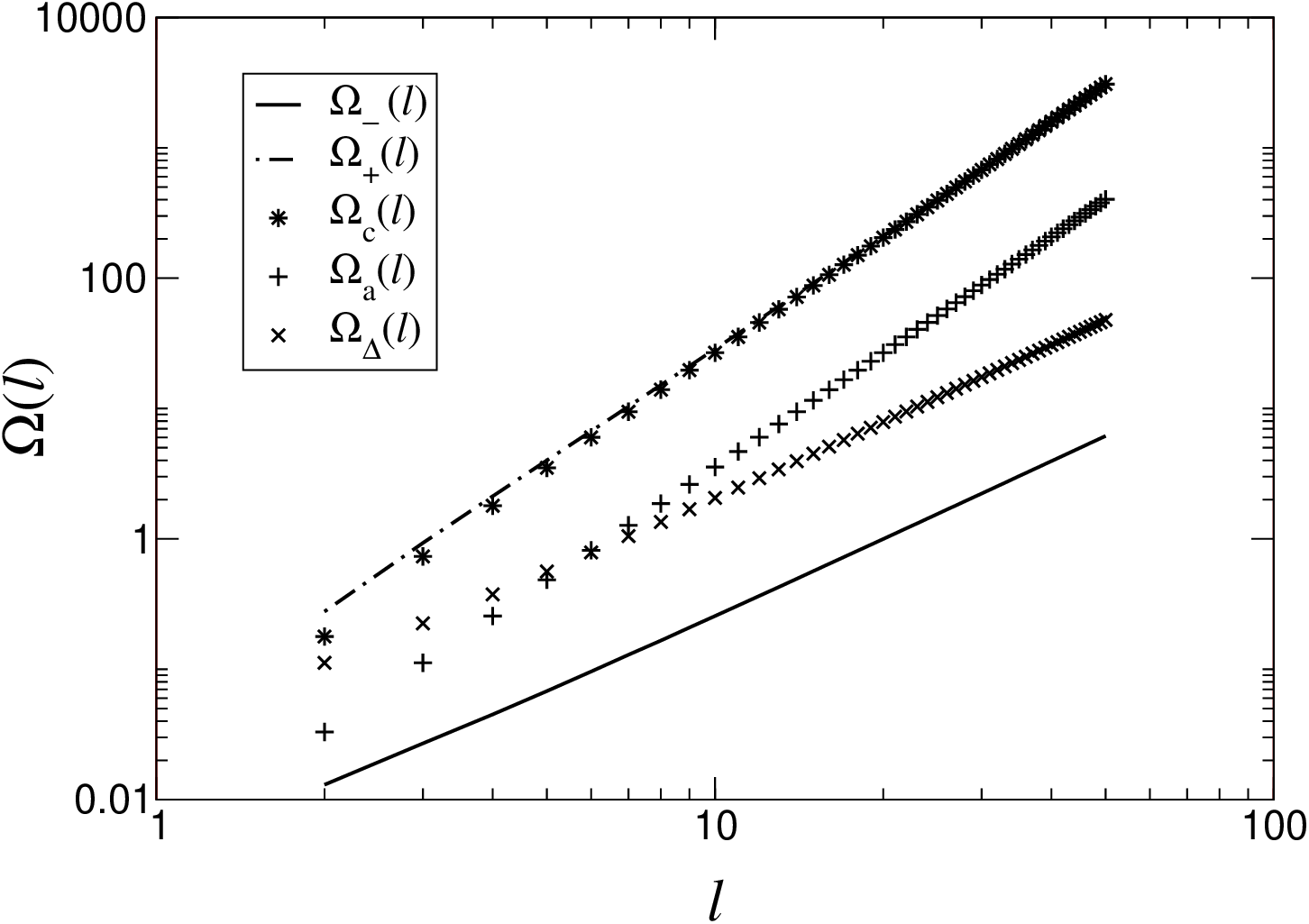}  
}}
\caption{The same dispersion relations as those illustrated in Fig.~2, for the same
parameter values but one: $\Sigma=0$. }
\end{figure}

For the sake of completeness of the discussion on the dissipative dynamics, we mention
an analytical relationship between $\Omega_{-}(\lmode)$ and $\Omega_{+}(\lmode)$, which
may be exploited in analysis of experimental data.  This relationship and its potential
have already been pointed out independently in Ref.~\cite{pott}.  It follows straightforwardly
from Eq.(\ref{dispersion}) that 
\begin{eqnarray}
\Omega_{-}(\lmode) \; \Omega_{+}(\lmode)
&=& \Omega_{\mathrm{a}}(\lmode) \; \Omega_{\Delta}(\lmode)  \nonumber \\
&=& \frac{\lmode (\lmode+1)(\lmode -1)(\lmode +2)\;\kappa_{\mathrm{a}}}
     {\eta R_0^3 \Gamma_{\lmode}} \left[ \lmode (\lmode +1) + \Sigma \right]\nonumber\\
&&\cdot   \frac{\lmode (\lmode+1)}{2R_{0}^{2}} \; \left( \frac{k_{\mathrm{eff}}}{b} \right) \;.
\end{eqnarray} 
In the above expression, all quantities may be considered known except the ratio
$k_{\mathrm{eff}}/b$.  Thus, for vesicle systems where the two relaxation processes
can both be resolved experimentally, the product $\Omega_{-}(\lmode) \; \Omega_{+}(\lmode)$,
and in turn, $k_{\mathrm{eff}}/b$, can be quantitatively determined.  This quantitative
information may be used to obtain a quantitative measure of the intermonolayer friction
coefficient $b$, provided that $k_{\mathrm{eff}}$ can be obtained by use of other independent
methods, e.g. the micromechanical technique \cite{rawicz}.  This method of determining $b$ can
serve both as an alternative to, and a check of, the method used previously for estimating
the value of $b$ \cite{evans92}. 

Finally, we now return to the issue of the validity of the no-inertia approximation,
which has already been mentioned in Section~4.  This discussion is relevant, especially
in view of the recent publication of Pott and M\'{e}l\'{e}ard~\cite{pott}, in which the experimental
data seems to suggest that the dynamics of quasi-spherical vesicles is not purely
dissipative, but that the relaxation is modulated by an oscillatory component of a small,
but observable amplitude.  Recapitulating Eq.~(\ref{Reynolds}), and using
$\rho_{\mathrm{b}}= 1\;\mathrm{g}/\mathrm{cm}^{3}$ and 
$\eta = 0.01\; \mathrm{erg}\cdot \mathrm{sec}/\mathrm{cm}^{3}$ for water, we have
\[
\bar{\mathcal{R}} = 10^{-6} \frac{\alpha^{2}}{t_{0}} \;.
\]
Setting $\alpha=60$ which corresponds to the lowest spherical harmonic mode, and using
the shortest time scales given in Table~1 and Table~2, for the two different cases
a) and b), respectively, we get two corresponding quantitative estimates,
$\bar{\mathcal{R}} \simeq 0.04$ and $\bar{\mathcal{R}} \simeq 0.001$.  The issue is
then really whether these numbers can be effectively treated as zero.  A theoretical
analysis that may address this issue clearly must take into account the inertial term
in the bulk hydrodynamics and is outside the scope of this paper.  But, if the
approximation turns out to be valid, then the theory we have presented here does not
provide an explanation for the observed oscillatory behaviour.  One may indeed question
whether the observed oscillatory behaviour is genuine.  If, however, the oscillatory
behaviour is genuine, then the no-inertia approximation should be re-examined by studying
the effect of the inertial term on the vesicle dynamics .  Moreover, one needs to bear
in mind that, if it is indeed the inertial term that is responsible for the oscillatory
behaviour, the time scales that characterize the part of the exponential relaxation
of the dynamics should in principle be qualitatively different from the times scales
obtained from a theory of purely dissipative dynamics such as the one presented in this
paper.  Looked at from this point of view, the interpretation given by the authors of
Ref.~\cite{pott} of their experimental data, which was based on their earlier theory of
purely dissipative dynamics \cite{bivas99}, does not appear consistent with the same data
which shows the presence of an oscillation.  

For the dynamics of the density-sum field, whose purely dissipative dynamics would be
characterized by a time scale of the order of $10^{-7}$ seconds, it is certain that 
the no-inertia approximation does not hold.  Consequently, the dynamics of this
field will contain an oscillatory component.  We may argue, however, that this
dynamics will have no observable effect on the dynamics of the other two modes.  One 
argument is that the coupling of the density-sum field to the other two fields implied
by the free energy is very weak for giant vesicles (and is in fact non-existent for
planar membranes, at least within the framework of linearized theories).  Another
argument is that the time scales associated with the relaxation-oscillation dynamics
of the density-sum field would still be very short compared to the time scales 
characterizing the dynamics of the other two fields. 
 
We would like to end this paper by stating our two main hopes that come with the
presentation of the work.  First, the current theoretical work will encourage more
systematic and careful experimental work that will lead to a clear and quantitative
understanding of the dynamics of vesicle shape fluctuations;  second, the fact that
the theory is built upon certain fundamental principles and considerations, together
with the systematic nature of its formulation, will be exploited in terms of extensions
of the theory that will describe vesicle systems formed of fluid bilayers containing
more than one molecular species (e.g. a second lipid or a protein).  

\vspace{3cm}
\noindent
{\small
{\bf {\sffamily Aknowledgement}} \\
LM and ML acknowledge the financial support from University of Southern Denmark
in the form of a research grant and in the form of a Ph.D. fellowship. The authors
are grateful to Per Lyngs Hansen, John Hjort Ipsen, Jonas Henriksen and Ole G. Mouritsen
for helpful discussions and critical reading of the manuscript.  The MEMPHYS Center for
Biomembrane Physics is supported by the Danish National Research Foundation.} 

\newpage
\begin{appendix}
\section{{\sffamily Appendix}}
In this Appendix, we will sketch briefly the calculations of static correlation
functions based on the free-energy model, Eq.(\ref{model2b}).  Eq.(\ref{model2b}) can
be expressed in a form more convenient for the calculations presented here:
\begin{eqnarray}
\label{app_model2b} 
\hat{F}&= \int d\hat{A} &\Big\{ k_{\mathrm{eff}} (\hat{\phi}^{\Delta})^2
                 + k_{\mathrm{eff}} (\hat{\phi}^{\Sigma})^2  
                 + 2 \kappa_{\mathrm{eff}} \hat{H}^2  \nonumber\\
&&\;\;               + 2 \lambda \hat{H} \hat{\phi}^{\Delta} +\sigma_{0} \Big\} \;,
\end{eqnarray}
where the use of the ``hat" over the variables indicates that they should be considered
as stochastic variables.

The calculations of the static correlation functions are essentially based on a
Gaussian theory of the fluctuations.  An inevitable issue in the calculations
concerns the choice of an appropriate set of {\em independent} degrees of freedom, 
whose fluctuations must be integrated over.  Although $\hat{u}_{\lmode m}$,
$\hat{\psi}^{\Delta}_{\lmode m}$ and $\hat{\psi}^{\Sigma}_{\lmode m}$, as defined in
Eqs.(\ref{u_expansion})-(\ref{phi_expansion2}), respectively, are independent of each other and
appear directly in the free-energy expression, they are not appropriate in general.  To make
the point clear, consider a local area element of membrane surface, $A_{\Delta}$, associated with
which are $N_{\Delta}^{+}$ and $N_{\Delta}^{-}$ lipid molecules in the outer and inner monolayer,
respectively.  Changes in the local monolayer density fields, $\phi^{\Delta}$ and $\phi^{\Sigma}$
can then be brought about by two types of independent physical processes: lateral flows of lipid
molecules at fixed surface geometry, which lead to changes in $N_{\Delta}^{+}$ and $N_{\Delta}^{-}$
at fixed $A_{\Delta}$; and, a change in the surface geometry at fixed $N_{\Delta}^{+}$ and
$N_{\Delta}^{-}$, which changes $A_{\Delta}$.   Thus, it is more meaningful to choose, together with
the $u-$field, the fluctuating fields that reflect the former process, as the appropriate set of
independent degrees of freedom.  Such fields, denoted henceforth by $\hat{n}^{\pm}(\theta,\varphi)$,
are related to the apparent density fields as follows:
\begin{eqnarray}
\hat{\rho}^{\pm}(\theta,\varphi) &\equiv &[\rho^{\pm}_{0} + \rho_{0} \hat{n}^{\pm}(\theta,\varphi) ]
                                   \frac{1}{\sqrt{g(\hat{u})}}\nonumber\\
&=& [\rho^{\pm}_{0} + \rho_{0} \hat{n}^{\pm}(\theta,\varphi) ]
                                [1 + g_{2}(\hat{u})] \;,
\end{eqnarray}  
where $g(\hat{u})\equiv a(\hat{u})/R_0^2$ and 
$g_{2}(\hat{u}) \equiv -2\hat{u} + 3\hat{u}^{2} - (\vec{\nabla}_{\mathrm{L}}\hat{u})^{2}/2$ 
is the expansion of $1/\sqrt{g(\hat{u})}-1$ up to second order in $\hat{u}$.  It follows easily
that 
\begin{eqnarray}
\hat{\phi}^{\Delta}(\theta,\varphi) &=& \phi_{0}^{\Delta} + \hat{n}^{\Delta}(\theta,\varphi) 
                                  + \phi_{0}^{\Delta}g_{2}(\hat{u}) 
                                  + \hat{n}^{\Delta}(\theta,\varphi)g_{2}(\hat{u}) \nonumber \\
\hat{\phi}^{\Sigma}(\theta,\varphi) &=& \hat{n}^{\Sigma}(\theta,\varphi) + g_{2}(\hat{u})  
                                  + \hat{n}^{\Sigma}(\theta,\varphi)g_{2}(\hat{u}) \;,                             
\end{eqnarray}
where $\hat{n}^{\Delta}(\theta,\varphi)
       \equiv (\hat{n}^{+}(\theta,\varphi) - \hat{n}^{-}(\theta,\varphi))/2$
and
$\hat{n}^{\Sigma}(\theta,\varphi)
       \equiv (\hat{n}^{+}(\theta,\varphi) + \hat{n}^{-}(\theta,\varphi))/2$.
       
Expanding the free energy given in Eq.(\ref{app_model2b}) to second order in the fluctuations
degrees of freedom, eliminating $\hat{u}_{00}$, $\hat{n}^\Sigma_{00}$ and $\hat{n}^\Delta_{00}$ using the fixed-volume and the fixed-molecular number constraints
\begin{eqnarray}
\hat{u}_{00}&=&-\frac{1}{\sqrt{4\pi}}\sum_{l\ge 1}\sum_{m=-l}^{l}\left|\hat{u}_{\lmode m}\right|\ ,\\
\hat{n}^\Delta_{00}&=&\mathrm{Const.}\ ,\\
\hat{n}^\Sigma_{00}&=&\mathrm{Const.}\ ,
\end{eqnarray}
leads to the following expression:             
\begin{eqnarray}
\hat{F} & = & \mathrm{Const.} \nonumber \\
    & & + \sum_{\lmode,m}^{\prime} \bigg\{ 
                     \frac{1}{2}\big\{ (\lmode -1)(\lmode +2)[\lmode (\lmode +1)\kappa_{\mathrm{eff}}
                         + \sigma_{0}R_{0}^{2} \nonumber\\
&&\hspace{.8cm}- \kappa_{\mathrm{s}}(\phi_{0}^{\Delta})^{2}
                         + 4 \lambda R_{0} \phi_{0}^{\Delta}]
                         + 8\kappa_{\mathrm{s}}[1+(\phi_{0}^{\Delta})^{2} ] \big\}  \;
                         |\hat{u}_{\lmode m}|^{2} \nonumber \\
                & & \hspace{.8cm} + \kappa_{\mathrm{s}}\;(\hat{n}^{\Delta}_{\lmode m})^{2}
                   + \kappa_{\mathrm{s}}\;(\hat{n}^{\Sigma}_{\lmode m})^{2}-  [(\lmode -1)(\lmode +2)\lambda R_{0} 
                                   \nonumber \\
                & & \hspace{.8cm}    + 4  \kappa_{\mathrm{s}} \phi_{0}^{\Delta} ] \;
                                    \hat{u}_{\lmode m}^{*} \hat{n}^{\Delta}_{\lmode m} 
                            - 4  \kappa_{\mathrm{s}} \; \hat{u}_{\lmode m}^{*} \hat{n}^{\Sigma}_{\lmode m} 
                         \bigg\} \;,  
\end{eqnarray}
where $\kappa_{\mathrm{s}} \equiv k_{\mathrm{eff}} R_{0}^{2}$.

Based on the above Gaussian theory for the fluctuations, and noting that, to the first
order in fluctuations, 
\begin{equation}
\hat{\psi}^{\Delta}_{\lmode m} = \hat{n}^{\Delta}_{\lmode m} - 2 \phi_{0}^{\Delta}\hat{u}_{\lmode m}
\;, 
\end{equation}
we finally obtain the following two static correlation functions,
\footnote{Other correlation functions can be obtained just as straightforwardly, but they are not listed here.}
\begin{equation}
\langle |\hat{u}_{\lmode m}|^{2} \rangle = \frac{k_{\mathrm B}T}
       {(\lmode -1)(\lmode +2)\kappa_{\mathrm a} [\lmode (\lmode + 1) + \Sigma ]} \;,
\end{equation}
which has been quoted in Eq.(\ref{static_u}) and Eq.(\ref{Sigma}), and
\begin{equation}
\langle \hat{u}_{\lmode m}^{*} \hat{\psi}^{\Delta}_{\lmode m} \rangle
= \frac{(\lmode -1)(\lmode +2)\lambda R_{0}}{2\kappa_{\mathrm{s}}} \langle |\hat{u}_{\lmode m}|^{2} \rangle 
\;.
\end{equation}

\end{appendix}

\noindent








\end{document}